\documentclass[final,12pt]{clear2024} 

\title[Inference of nonlinear causal effects]{Inference of nonlinear causal effects with GWAS summary data}
\usepackage{times,booktabs,bm,mwe,xcolor}
\usepackage[linesnumbered,ruled]{algorithm2e}
\newtheorem{condition}{Condition}[section]

\newcommand{\E}{\operatorname{E}} 
\newcommand{\I}{\operatorname{I}} 

\newcommand{\Cov}{\operatorname{Cov}}

\newcommand{\independent}
{\mathrel{\text{\scalebox{1.07}{$\perp\mkern-10mu\perp$}}}}

\newcommand{\argmin}{\operatorname*{\arg\min}}
\newcommand{\argmax}{\operatorname*{\arg\max}}

\newcommand{\dto}{\overset{d}{\longrightarrow}}

\newcommand{\sign}{\operatorname{sign}} 

\clearauthor{%
 \Name{Ben Dai}\thanks{Equal contribution.} \Email{bendai@cuhk.edu.hk}\\
 \addr Department of Statistics, The Chinese University of Hong Kong
 \AND
 \Name{Chunlin Li}$^*$ \Email{li000007@umn.edu}\\
 \addr School of Statistics, The University of Minnesota
 \AND
 \Name{Haoran Xue} \Email{xuexx268@umn.edu}\\
 \addr Division of Biostatistics, The University of Minnesota
 \AND
 \Name{Wei Pan} \Email{panxx014@umn.edu}\\
 \addr Division of Biostatistics, The University of Minnesota
 \AND
 \Name{Xiaotong Shen} \Email{xshen@umn.edu}\\
 \addr School of Statistics, The University of Minnesota
}

\begin{document}

\maketitle

\begin{abstract}%
  Large-scale genome-wide association studies (GWAS) have offered an exciting opportunity to discover putative causal genes or risk factors associated with diseases by using SNPs as instrumental variables (IVs). However, conventional approaches assume linear causal relations partly for simplicity and partly for the availability of GWAS summary data. 
  In this work, we propose a novel model {for transcriptome-wide association studies (TWAS)} to incorporate nonlinear relationships across IVs, an exposure/gene, and an outcome, which is robust against violations of the valid IV assumptions, permits the use of GWAS summary data, and covers two-stage least squares as a special case.  We decouple the estimation of a marginal causal effect and a nonlinear transformation, where the former is estimated via sliced inverse regression and a sparse instrumental variable regression, and the latter is estimated by a ratio-adjusted inverse regression. 
  On this ground, we propose an inferential procedure. An application of the proposed method to the ADNI gene expression data and the IGAP GWAS summary data identifies 18 causal genes associated with Alzheimer's disease, including APOE and TOMM40, in addition to 7 other genes missed by two-stage least squares considering only linear relationships.
  Our findings suggest that nonlinear modeling is required to unleash the power of IV regression for identifying potentially nonlinear gene-trait associations. Accompanying this paper is our Python library 
  \texttt{nl-causal} (\url{https://nonlinear-causal.readthedocs.io/}) 
  that implements the proposed method.
\end{abstract}

\begin{keywords}%
  nonlinear causal effect; privacy-constrained datasets; sliced inverse regression; genome-wide association study; transcriptome-wide association study; two-sample inference; 
\end{keywords}

\section{Introduction}

Causal inference methods in transcriptome-wide association studies (TWAS) have successfully discovered numerous (putative) \textit{causal genes} associated with complex traits and diseases \citep{Gusev2016ng}, using genetic variants, typically single nucleotide polymorphisms (SNPs), as instrumental variables (IVs) \citep{Yang2010ng}. Understanding these gene-to-disease associations has considerable ramifications in the field of genomics, possibly spearheading a much-anticipated revolution in personalized and precision medicine.
\smallbreak

\noindent \textbf{2SLS in TWAS.} Conventional TWAS applies two-sample two-stage least squares (2SLS; \citet{kang2016instrumental}) to integrate expression quantitative trait locus (eQTL) data for gene expression and genome-wide association study (GWAS) summary data for a trait of interest,  thereby pinpointing potential causal genes for disease risk, such as Alzheimer's Disease (AD).
Specifically, we denote instrumental variables as $\bm z\in\mathbb{R}^p$, a scalar exposure as $x\in\mathbb{R}$, and a scalar outcome as $y\in\mathbb{R}$. For example, SNPs ($\bm{z}$) are used as instrumental variables for a gene's expression ($x$) to identify its causal association with AD risk ($y$). 2SLS assumes that $(\bm z, x, y)$ satisfy a two-stage \textit{linear} model:
\begin{equation}
    \label{model:2SLS}
    x = \bm z^\intercal \bm\theta + w, \qquad y = \beta x + \bm z^\intercal \bm\alpha + \varepsilon, 
\end{equation}
where $(w,\varepsilon)$ are the error terms independent of the instruments $\bm z$, however, $w$ and $\varepsilon$ may be correlated due to underlying confounders, and $\beta\in\mathbb{R}$, $\bm\alpha\in\mathbb{R}^p$, $\bm\theta\in\mathbb{R}^p$ are unknown parameters. 

The primary objective of 2SLS is for statistical inference on the causal effect $\beta$ of the exposure $x$ on the outcome $y$ based on \eqref{model:2SLS}. The estimation of $\beta$ via 2SLS can be executed in two stages: (Stage 1) 2SLS utilizes IVs $\bm{z}$ to predict the exposure $x$ via linear regression, subsequently providing an estimate $\widehat{\bm{\theta}}$; (Stage 2) the estimated ``debiased'' exposure (obtained as $\widehat{x} = \widehat{\bm{\theta}}^\intercal \bm{z}$) is used to estimate the causal effect $\beta$ via a regression from $\widehat{x}$ to $y$. Consequently, 2SLS produces \textit{unbiased} estimation of the causal effect from exposure to the outcome by mitigating confounder-induced bias. 
Another key benefit of 2SLS is its ability to infer based solely on the summary statistics of $x$-$\bm{z}$ and $y$-$\bm{z}$ correlations. This feature is particularly beneficial for \textit{privacy-constrained} datasets, such as SNP genotype data.
In the content of TWAS, for each gene being treated as an exposure, 2SLS first builds a predictive model using its cis-SNPs around this gene as IVs for the expression level with the eQTL data. Then the predicted gene expression is obtained with the GWAS summary data and tested for association with the trait to determine whether the gene is putatively causal to the trait.

\begin{figure}[h]
    \centering
    \includegraphics[scale=.33]{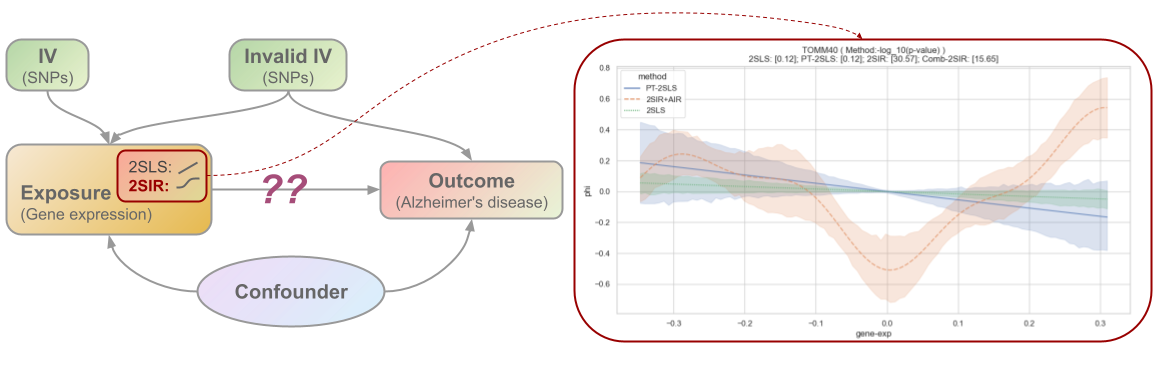}
    \caption{\textbf{Left.} {A structure plot of the proposed 2SIR model, which admits a nonlinear causal effect from exposure to outcome.} \textbf{Right.} Estimated transformations of TOMM40 (a well-known AD gene) based on {2SLS}, {PT-2SLS}, and our {2SIR+AIR}, and the resulting p-values are included in the title, yielding that TOMM40 is only identified by our method.
    Moreover, the $R^2$s for the stage one model on $\widehat{\phi}(x) \sim \widehat{\bm{\theta}}^\intercal \bm{z}$ are 0.230 (2SLS), 0.230 (PT-2SLS), and 0.253 (2SIR), suggesting that the nonlinear model \eqref{model:structural-equation} is suited for this data.}
    \label{fig:app_demo}
    \vspace*{-.4cm}
\end{figure}

Despite the substantial advantages of the TWAS using 2SLS in causal inference, a primary limitation surfaces due to its inherent assumption of linearity. Previous TWAS studies \citep{gamazon2015gene,Gusev2016ng,zhu2016integration} generally propose a linear relationship between cis-SNPs and gene expression in the first stage and between gene expression and a GWAS trait/outcome in the subsequent stage. This framework overlooks the likely existence of nonlinear effects \citep{Mackay2014nrg}.
On the other hand, to our knowledge, none of the existing non-parametric IV regression methods are applicable to GWAS summary data,  while individual-level GWAS data are usually unavailable due to privacy and logistic issues, presenting challenges to incorporating flexible nonlinear models into TWAS with GWAS summary data. In our motivating example, the individual-level AD GWAS data from many sub-studies are unavailable, but its meta-analyzed summary data are available.
Some recently proposed methods \citep{zhang2017use,okoro2021transcriptome}
relax the linear assumption in stage 1, 
while others do so in stage 2 \citep{he2023delivr},
which however requires the use of individual-level data.
Misspecification of a nonlinear effect as a linear (or other specific) one may distort subsequent causal inference, damping the statistical power of the TWAS method. For illustration, we consider the eQTL data for a well-known AD-related gene, TOMM40, from our real data example;
see Section \ref{sec:app} for more details. {Figure \ref{fig:app_demo} provides
some compelling evidence for the nonlinear effects in both stages of TWAS. In the first stage, it displays a nonlinear relationship between the cis-SNPs and the gene expression level of TOMM40, as evidenced by a higher $R^2$ value of the nonlinear model over those of its linear competitors. In the second stage, a nonlinear causal association of TOMM40 with the AD risk is strongly corroborated by the highly significant p-value obtained with our method.
Consequently, this well-known AD gene is successfully identified by our proposed method (2SIR+AIR) but missed by both 2SLS and its power-transformed extension (PT-2SLS), suggesting the necessity of nonlinear modeling in TWAS.}

{Moreover, as an IV regression method, conventional TWAS relies on three key IV assumptions to remove the hidden confounding effects: (IV1) the IVs are associated with the exposure, (IV2) the IVs are not associated with the outcome conditional on the exposure, and (IV3) the IVs are not associated with the unmeasured confounders conditional on the exposure. While (IV1) is straightforward to handle, (IV2) and (IV3) are fragile in practice due to the widespread pleiotropy of SNPs \citep{solovieff2013pleiotropy}. 
This phenomenon refers to the situation when an SNP affects the GWAS trait/disease not mediated through exposure, violating (IV2) and/or (IV3) and causing severe bias in causal inference.}
A line of recent works \citep{kang2016JASA,windmeijer2019JASA,guo2018JRSSB} has been focusing on the violation of (IV2) and/or (IV3). Of note, these methods use linear models, and their nonlinear counterparts remain unexplored.

\smallbreak

\noindent \textbf{Other methods.}
Besides TWAS, Mendelian Randomization (MR) is another important and popular subject in genetics that uses SNPs as IVs to infer a causal relationship between an exposure and an outcome, typically two complex traits \citep{morrison2020mendelian,xue2021constrained}.
Both TWAS and conventional MR are two-stage IV regression methods for causal inference, and they share many similarities, yet their implementations are different due to distinct types of data being used. 
Although both TWAS and conventional MR use GWAS summary data in the second stage, in the first stage MR uses GWAS summary data of sample size typically in tens of thousands or even larger, while TWAS typically uses individual-level eQTL data of sample size in a few hundreds or at most one or two thousands. 
Usually, SNPs being used in TWAS are around the target gene (i.e. cis-SNPs) and are correlated, while most MR methods use independent SNPs from the whole genome. Due to these distinctions, the existing typical MR methods do not fit the TWAS analysis.

In a nutshell, nonlinear modeling that is robust to the violation of IV assumptions and at the same time leverages large-scale GWAS summary data lacks for TWAS analysis.
To addressing the limitations of existing methods, we develop an approach with the following novel aspects.
\begin{itemize}
  \setlength\itemsep{0em}
  \item We propose a flexible model to admit an \emph{arbitrary unknown nonlinear} causal relationship between an exposure and an outcome. Importantly, the proposed model is applicable to GWAS summary data while being robust to invalid IVs, and \emph{covers 2SLS} as a special case.

  \item Based on the proposed model, we decouple the estimation of a  \emph{causal effect}
  and a \emph{nonlinear causal transformation}. The inference of the causal effect are established by the proposed 2SIR based on sliced inverse regression. 
  Then, the unknown nonlinear transformation can be estimated by the proposed AIR. The validity of the proposed hypothesis testing and interval estimation is ensured by our theoretical result, and verified by extensive simulation study.
  


  \item The ADNI data and the IGAP GWAS summary data confirm the efficacy of our approach.
  The results (Section \ref{sec:app}) indicate that our method successfully replicates the significant AD genes identified by 2SLS, while uniquely identifying 7 additional causal genes.
  Our real data analysis suggests that
  nonlinear modeling is suited to unleash the power of TWAS.
\end{itemize}

\section{Nonlinear modeling of TWAS data} 

We denote a vector of IVs as $\bm z\in\mathbb{R}^p$, a scalar exposure as $x\in\mathbb{R}$, and a scalar outcome as $y\in\mathbb{R}$. 
In our TWAS case study (cf. Section \ref{sec:app}), SNPs are used as instrumental variables for a gene's expression to identify its causal association with the AD risk.
Without loss of generality, we assume $(\bm z, x, y)$ has mean zero. 
Suppose $(\bm z, x, y)$ satisfy a nonlinear model 
\begin{equation}
    \label{model:structural-equation}
    \phi(x) = \bm z^\intercal \bm\theta + w, \qquad y = \beta \phi(x) + \bm z^\intercal \bm\alpha + \varepsilon, 
\end{equation}
where $(w,\varepsilon)$ are the error terms independent of the instruments $\bm z$,
and $\beta\in\mathbb{R}$, $\bm\alpha\in\mathbb{R}^p$, $\bm\theta\in\mathbb{R}^p$ are unknown parameters, 
and $\phi(\cdot)$ is an unknown transformation.

The following provides some in-depth motivations for the proposed model \eqref{model:structural-equation}.
First, as shown by others \citep{lin2022accounting,he2023delivr} and to be shown here, there is empirical evidence to support the existence of non-linear effects that certain genes have on various traits, thus the possible non-linear function $\phi(x)$ in \eqref{model:structural-equation}.
Second, it is well known that, due to the small effect sizes of SNPs on complex traits, linear models for the effects of SNPs perform well in practice, hence we adopt the widely-used linearity assumption of  $\bm z$, which (implicitly) connects the two-stage models in \eqref{model:structural-equation}.
An alternative, and perhaps more popular, non-linear model as used in \citet{hartford2017deep,he2023delivr} would be a linear model of the effects of SNPs $\bm{z}$ on the gene expression $x$ in Stage 1 but a similar non-linear Stage 2 model as in \eqref{model:structural-equation}, which however would imply a non-linear model for the effects of SNPs $\bm{z}$ on trait $y$.
This perhaps is debatable: since the causal pathway is likely to be from SNPs $\bm{z}$ to gene $x$ then to trait $y$, the effect sizes of SNPs (i.e. their heritabilities) are expected to be smaller on $y$ than on $x$, suggesting that if a linear model of $\bm{z}$ on $x$ is reasonable, another linear model of $\bm{z}$ on $y$ should approximately hold. 
In fact, it was shown empirically that, even if a linear model of the effects of SNPs $\bm{z}$ on a gene's expression level $x$ was reasonable in Stage 1, assuming a linear model of $\bm{z}$ on $x^2$ (Stage 1 in our model) performed better than a non-linear model (as implied by the linearity of $\bm{z}$ on $x$), again likely due to the small effect sizes of SNPs and the parsimony of linear models (see Remarks subsection in Materials and Methods section of \citet{lin2022accounting}). 
Importantly, the implicit linear structure allows the use of GWAS summary data of our method, in contrast to requiring individual-level data by the other non-linear models.


Furthermore, our model \eqref{model:structural-equation} holds two significant advantages over 2SLS \eqref{model:2SLS}.
First, the assumptions of \eqref{model:structural-equation} are weaker than the classical 2SLS. Specifically, \eqref{model:structural-equation} admits an \emph{arbitrary} nonlinear transformation $\phi(\cdot)$ across $\bm{z}$, $x$ and $y$, relaxing the linearity assumption in the standard TWAS/2SLS. Second, it includes 2SLS and Yeo-Johnson power transformation 2SLS (PT-2SLS) \citep{yeo2000new} as special cases. 
It is worth mentioning that the proposed method remains competitive against 2SLS/PT-2SLS even if the linear 
assumption or normality assumption holds; see Section \ref{sec:sim}.
Overall, the proposed model \eqref{model:structural-equation} is a natural extension of 2SLS.

In \eqref{model:structural-equation}, $\beta\phi(\cdot)$ represents the 
influence of the exposure on the outcome, which is our primary focus,
while $\bm\alpha$ and $\bm\theta$ are nuisance parameters. 
In particular, $\bm\alpha\neq \bm 0$ indicates the violation of the second and/or third IV assumptions. 
Generally, the effect $\beta\phi(\cdot)$ may not be identifiable with the presence of invalid IVs. 
In the literature, additional structural constraints are imposed to avoid this issue.  
For example, if $\|\bm\alpha\|_0 < p/2$ is known a priori, then $\beta\phi(\cdot)$ becomes well-defined \citep{kang2016instrumental}. 
Furthermore, 
note that $\beta$ and $\phi$ are only identifiable up to a multiplicative scalar,
even if $\beta\phi(\cdot)$ is well-defined in \eqref{model:structural-equation}.
Thus, we fix $\|\bm\theta\|_2 = 1$ and $\beta \geq 0$ in the subsequent discussion so that $\beta$ and $\phi$ are identifiable. 


On this ground, Definition \ref{def:causal_effect} summarizes the quantities of interest.
\begin{definition}[Causal effect and transformation]
    \label{def:causal_effect}
    In \eqref{model:structural-equation}, let $\|\bm{\theta}\|_2 =1$ and $\beta \geq 0$. Then, \\
    \indent    (i) $\beta$ is called the marginal causal effect; \\
    \indent    (ii) $\phi(\cdot)$ is called the nonlinear transformation (of the exposure); \\
    \indent    (iii) $\beta \phi(\cdot)$ is called the nonlinear effect function.
\end{definition}

Specifically, $\beta$ summarizes the marginal effect of the causal influence of the exposure $x$ on the outcome $y$, 
in that $\beta > 0$ indicates the presence of the causal relation, and the corresponding hypothesis testing and confidence interval are developed in Sections \ref{sec:est_beta}. 
It is worth noting that $\beta$ in \eqref{model:structural-equation} only represents the magnitude of the causal effect, which does not imply a positive/negative relation as in 2SLS, due to the nonlinear transformation $\phi(\cdot)$.
{If the model \eqref{model:structural-equation} is well-specified,
the nonlinear effect function $\beta \phi(\cdot)$ in (iii) can be used to measure the average treatment effect (ATE) between two exposure/treatment levels.} 
In our case study, $\beta>0$ indicates the presence of the causal influence of a gene on the AD risk, {and if the model \eqref{model:structural-equation} is well-specified, $\phi(\cdot)$ represents the potentially nonlinear pattern of a putative causal association}. 

Let $(\bm Z_{\nu},\bm X_{\nu},\bm Y_{\nu})$ be $n_{\nu}\times(p+2)$ matrix,
where each row $(\bm z_{\nu i},x_{\nu i},y_{\nu i})$, $1\leq i\leq n_{\nu}$, $\nu=1,2$, represents an independent observation from \eqref{model:structural-equation}. 
In what follows, assume that we have two independent samples 
$\mathcal D_1=\{\bm Z_1,\bm X_1\}$
and $\mathcal D_2=\{n_2^{-1}\bm Z_2^\intercal\bm Z_2, n_2^{-1}\bm Z_2^\intercal \bm Y_2, n_2^{-1}\bm Y_2^\intercal\bm Y_2\}$
from \eqref{model:structural-equation}.
Without loss of generality, we assume that $\bm{Y}_2$ is pre-normalized as $n_2^{-1} \bm{Y}^\intercal_2 \bm{Y}_2 = 1$.
Importantly, we require neither that all variables $(\bm z,x,y)$ are observed simultaneously, 
nor the availability of individual-level data $(\bm Z_2,\bm X_2,\bm Y_2)$,
allowing the application to summary statistics, 
{like GWAS summary data, for the second sample.}
Our goal is to infer $\beta$ and $\beta \phi(\cdot)$ from the observed data $\mathcal D_1,\mathcal D_2$. 
In the sequel, we propose estimating the marginal causal effect $\beta$ and the nonlinear transformation $\phi$ separately.

\subsection{Estimation and inference of marginal causal effect}
\label{sec:est_beta}

The proposed procedure for estimating $\beta$ consists of two stages. 
In the first stage, note that $x \independent \bm{z} \mid \bm{z}^\intercal\bm{\theta}$ in \eqref{model:structural-equation},
which coincides with a single index model \citep{duan1991slicing,cook2009regression}, 
and the sliced inverse regression (SIR; \citet{li1991sliced}) can be used to estimate $\bm \theta$.
Specifically, given the dataset $\mathcal D_1$, SIR divides the range of $x_i$ into $S$ non-overlapping slices $\text{Slice}_s (s = 1, \cdots, S)$,
and estimates ${\bm \theta}$ as the eigenvector of $\widehat{\bm{\Sigma}}^{-1} \widehat{\bm{\Gamma}}$ associated with the largest eigenvalue:
\begin{equation}
    \label{eqn:sir}
    \widehat{\bm \theta} = \argmax_{\bm \theta \in \mathbb{R}^p} \bm{\theta}^\intercal \widehat{\bm{\Gamma}} \bm{\theta}, \quad \text{s.t.} \quad \bm{\theta}^\intercal \widehat{\bm{\Sigma}} \bm{\theta} = 1, \quad \text{where } \widehat{\bm{\Gamma}} = \sum_{s=1}^S \frac{n_{1s}}{n_1} \bar{\bm{z}}_{(s)} \bar{\bm{z}}^\intercal_{(s)}, \ \bar{\bm{z}}_{(s)} = \frac{1}{n_{1s}} \sum_{x_i \in \text{Slice}_s} \bm{z}_{1i},
\end{equation}
where $\widehat{\bm{\Sigma}}$ is the sample covariance matrix of $\bm{z}$, and $\widehat{\bm{\Gamma}}$ is the between slice covariance matrix, 
with $n_{1s}$ being the number of samples in the $s$-th slice $\text{Slice}_s$.



In the second stage, we estimate $\beta$ via a sparse instrumental variable regression using the data $\mathcal D_2$. 
Specifically, note that the second equation in \eqref{model:structural-equation} can be rewritten as  
\begin{equation}\label{model:y-z}
    y = \bm z^\intercal\bm\theta\beta + \bm z^\intercal\bm\alpha + e, \qquad e = w\beta + \varepsilon,\qquad \E(e) = 0, \qquad \E (e^2) = \sigma_e^2.
\end{equation}
{Recall that $\alpha_j\neq 0$ indicate $z_j$ violates (IV2) and/or (IV3). Motivated by \citet{xue2021constrained}, we separate the potential bias due to invalid IVs from the causal effect $\beta$ via a sparse regression:
\begin{equation}\label{stage2:sparse-regression}
    \begin{split}
        \min_{\bm\alpha,\beta} \ 
        (\widehat{\bm\theta}\beta + \bm\alpha)^\intercal\bm Z_2^\intercal \bm Z_2 (\widehat{\bm\theta}\beta + \bm\alpha) 
        - 2\bm Y_2^\intercal\bm Z_2 (\widehat{\bm\theta}\beta + \bm\alpha)
        \quad \text{s.t.}\quad 
       \|\bm\alpha\|_0 \leq K,
    \end{split}
\end{equation}
where $\|\bm\alpha\|_0=\sum_{j=1}^p\I(\alpha_j\neq0)$ and $K\geq 0$ is an integer tuning parameter indicating the number of invalid IVs.} For implementation, $\|\cdot\|_0$ penalty can be replaced by a sparsity-inducing surrogate penalty, such as SCAD \citep{fan2001variable}, TLP \citep{shen2010grouping}, and MCP \citep{zhang2010nearly}. {In our data analysis, we use the SCAD as a computational surrogate; see Appendix \ref{sec:computation} for details.}  


Taken together, the proposed procedure consists of the estimation of $\bm\theta$ via a Sliced Inverse 
Regression, and that of $\beta$ via a Sparse Instrumental Regression. This methodology is named  
Two-Stage Instrumental Regression (2SIR), as summarized in Algorithm \ref{algo:2sir}.

\begin{algorithm}
    \SetKwInOut{Input}{Input}
    \Input{Datasets $\mathcal D_1$ and $\mathcal D_2$}
    (Stage 1: Sliced inverse regression) Estimate $\widehat{\bm{\theta}}$ via \eqref{eqn:sir} with $\mathcal D_1$ \;
    (Stage 2: Sparse instrumental regression) Estimate $\widehat{\beta}$ via \eqref{stage2:sparse-regression} with $\mathcal D_2$ and $\widehat{\bm{\theta}}$\;
    (Sign adjustment for identifiability) $\widehat{\bm\theta}\leftarrow \sign(\widehat{\beta})\widehat{\bm\theta}$,\qquad  $\widehat{\beta} \leftarrow |\widehat{\beta}|$\;
  \Return{Estimated causal effect $(\widehat{\beta},\widehat{\bm\theta})$}
    \caption{Two-stage instrumental regression (2SIR) for $\beta$ estimation}
    \label{algo:2sir}
\end{algorithm}




Next, we turn to present inferential procedures for the marginal causal effect $\beta$, including hypothesis testing and confidence intervals.
Before proceeding, Theorem \ref{theorem:2sir} summarizes the asymptotic properties of the 2SIR estimator.
\begin{theorem}\label{theorem:2sir}
    {Let $\widehat{\beta}$ be the 2SIR estimator produced by Algorithm \ref{algo:2sir} with $\|\cdot\|_0$ penalty, SCAD, TLP, or MCP being used in \eqref{stage2:sparse-regression}.}
    Assume Conditions \ref{condition:sir} and \ref{condition:sparse-regression} in Appendix \ref{sec:regularity-condition}. If $K=|A|$ in \eqref{stage2:sparse-regression} and $(w,\varepsilon)$ is normally distributed, then 
    \begin{align}
        &n_2^{1/2}(\widehat{\beta}-\beta) = |n_2^{1/2}\beta + \zeta - \eta| - n_2^{1/2}\beta + o_p(1),\quad
        \zeta\independent\eta, \nonumber \\
        &\zeta \sim N(0,\Omega_X {\sigma^2_{e}}), \quad 
        \eta \sim \sqrt{r}\beta \Omega_X \bm\theta^\intercal \widetilde{\bm \Sigma} \bm\xi, \nonumber
    \end{align}
    where $A = \{ j : \alpha_j \neq 0 \}$,
    $n_2/n_1 \to r$ and $n_1^{1/2}(\widehat{\bm\theta} -\bm \theta) \dto \bm\xi$,
    $\widetilde{\bm \Sigma} = \bm \Sigma 
    - \bm \Sigma_{*A}\bm \Sigma_{AA}^{-1} \bm \Sigma_{A*}$, 
    $\Omega_X=(\bm\theta^\intercal \widetilde{\bm \Sigma} \bm\theta )^{-1}$, and 
    $\bm \Sigma_{*A},\bm \Sigma_{A*}$ denote the columns and rows of $\bm \Sigma$ indexed by $A$, respectively. 
\end{theorem}

In Theorem \ref{theorem:2sir}, $(w,\varepsilon)$ is assumed to be normally distributed for simplicity, which is not critical to large-sample inference. Now, we infer $\beta$ based on Theorem \ref{theorem:2sir}. 
First, consider the hypotheses:
$
    H_0: \beta = 0 \ \text{versus} \ H_a: \beta > 0,
$
where rejecting the null hypothesis $H_0$ indicates evidence for causal influence of the exposure $x$ on the outcome $y$. Define the pivotal test statistic 
\begin{equation}\label{eqn:test-statistic}
        \widehat T = \frac{n_2^{1/2}\widehat\beta}{\widehat\sigma_e (\widehat{\bm\theta}^\intercal\widehat{\bm \Sigma}\widehat{\bm\theta} 
    - \widehat{\bm\theta}^\intercal \widehat{\bm\Sigma}_{*A}(\widehat{\bm\Sigma}_{AA})^{-1}\widehat{\bm\Sigma}_{A*}\widehat{\bm\theta} )^{1/2}}.
\end{equation}
Given a significance level $\alpha\in(0,1)$, the null hypothesis $H_0$ is rejected 
if and only if $\widehat T > \Phi^{-1}_{N(0,1)}(1-\alpha/2)$, where $\Phi^{-1}_{N(0,1)}(\cdot)$ denotes the quantile function of $N(0,1)$. As a consequence of Theorem \ref{theorem:2sir}, 
Corollary \ref{corollary:test} justifies the proposed test. 
\begin{corollary}\label{corollary:test}
    Assume the conditions in Theorem \ref{theorem:2sir}. The following statements are true. \\
    \indent (i) Under the null hypothesis $H_0: \beta=0$, we have 
    \begin{equation}\label{eqn:test-size}
        \limsup_{n_2\to\infty} P_{H_0}\Big( \widehat T > \Phi^{-1}_{N(0,1)}( 1-\alpha/2)  \Big)\leq \alpha.
    \end{equation}
    \indent (ii) Under the alternative hypothesis $H_a:\beta = n_2^{-1/2}h$, we have 
    \begin{equation*}
        \liminf_{n_2\to\infty} P_{H_a}\Big( \widehat T > \Phi^{-1}_{N(0,1)}( 1-\frac{\alpha}{2}) \Big)\geq P\left(|N( \Omega_X^{-1/2} \sigma_{e}^{-1} h ,1)|> \Phi^{-1}_{N(0,1)}( 1-\frac{\alpha}{2} )\right).
    \end{equation*}
\end{corollary}

Empirically, Section \ref{sec:sim} shows that the proposed test can control the Type I error under the null hypothesis $H_0$ while possessing desirable power under $H_a$. Moreover, we developed a combined test over a different number of slices for 2SIR, see Appendix \ref{sec:comb}.

Next, we consider constructing a valid CI for $\beta$.  
Indeed, this can be challenging,
since the asymptotics of the SIR estimator depends on an unknown distribution $\bm z\mid x$ 
\citep{zhu1995asymptotics}, which is intractable. 
To overcome this difficulty, we propose a resampling-based CI in light of Theorem \ref{theorem:2sir}. 
Specifically, by the triangle inequality, $n_2^{1/2}|\widehat{\beta} - \beta| \leq |\zeta - \eta| + o_p(1)$. Therefore, the CI of $\beta$ can be produced by resampling $|\zeta - \eta|$.

For implementation, we first compute $(\widehat{\bm\theta},\widehat\beta)$ via Algorithm \ref{algo:2sir}, denote $\widehat{\bm{\Sigma}}_R = \widehat{\bm\Sigma} - \widehat{\bm\Sigma}_{*A}\widehat{\bm\Sigma}_{AA}^{-1}\widehat{\bm\Sigma}_{A*}$, $\widehat\Omega_X = ( \widehat{\bm\theta}^\intercal \widehat{\bm{\Sigma}}_R \widehat{\bm\theta})^{-1}$, and $\widehat{\sigma}_e^2 = n_2^{-1}(\bm Y_2^\intercal\bm Y_2 - \bm Y_2^\intercal\bm Z_2(\bm Z_2\bm Z_2)^{-1}\bm Z_2^\intercal\bm Y_2)$. 
Then the bootstrap estimates $\widehat{\bm\theta}^*_l$s are computed as $\widehat{\bm\theta}^*_l = \sign(\widehat{\bm \theta}^\intercal \widetilde{\bm\theta}^*_l ) \widetilde{\bm\theta}^*_l $, where $\widetilde{\bm\theta}^*_l$ is computed via Step 1 (SIR) in  Algorithm \ref{algo:2sir} based on resampling $\mathcal D_1$, and $\zeta^*_l\sim N(0,\widehat\Omega_X\widehat\sigma_e^2)$ is generated according to its asymptotic distribution; $l=1,\ldots,M$, where $M$ is the Monte-Carlo size.
In this way, we approximate the distribution of $\eta$ by the Monte-Carlo sample: for $l=1,\ldots,M,$
$
\eta^*_l =  \frac{1}{2} n_2^{1/2} \widehat\beta \widehat\Omega_X \big( (\widehat{\bm\theta_l^*})^\intercal \widehat{\bm{\Sigma}}_R \widehat{\bm\theta}^*_l - \widehat{\bm\theta}^\intercal \widehat{\bm{\Sigma}}_R \widehat{\bm\theta} \big).
$
Hence, the $(1-\alpha)$-confidence interval is: 
\begin{equation}\label{eqn:confidence-interval}
    \beta \in \left[ \max(0,\widehat\beta - n_2^{-1/2} \widehat Q^*(1-\alpha)),\ \widehat\beta + n_2^{-1/2} \widehat Q^*(1-\alpha) \right],
\end{equation}
where $\widehat Q^*(\cdot)$ is the quantile function of {$(|\zeta^*_l - \eta_l^*|)_{l=1}^M $}.
Finally, for interval estimation of $\beta\phi(\cdot)$, there is ample literature devoted to constructing nonparametric confidence bands; see \citep{hall2013simple}.

Section \ref{sec:sim} indicates that the proposed method yields peak performance in the estimation and inference of the marginal causal effect in various simulated examples. 
Yet, in practice, visualization of $\phi$ may shed light on the specific relationship between the exposure and outcome.
In the next section, we develop an algorithm to estimate the nonlinear transformation $\phi$.

\subsection{Estimation of nonlinear transformation}

The challenge of estimating $\phi(\cdot)$ is twofold. 
First, individual-level data of $(z,x,y)$ are usually unavailable, 
preventing the estimation of $\phi$ from the second equation of \eqref{model:structural-equation}. 
Second, $w$ is correlated with $\phi(x)$ in \eqref{model:structural-equation}, 
rendering a biased estimator when for example a least-squares regression of $\bm{z}^\intercal \bm{\theta}$ is conducted over $x$. To address these issues, we propose an Adjusted Inverse Regression (AIR) for consistent estimation of $\phi$. 
An important observation is made in Proposition \ref{prop:AIR}, 
showing that the transformation $\phi$ is proportional to the least-squares estimator.

\begin{proposition}
    \label{prop:AIR}
    Suppose $\E(\bm z^\intercal \bm\theta \mid x) = \E(\bm z^\intercal \bm\theta \mid \phi(x))$ and $(\bm z^\intercal\bm\theta,w)$ has an elliptically symmetric distribution. 
    Then there exists a constant $\rho$ such that $\phi(x) = \rho\E(\bm z^\intercal\bm\theta\mid x)$. 
\end{proposition}

In light of Proposition \ref{prop:AIR}, $\phi$ can be estimated by a two-stage procedure. 
First, we estimate the conditional mean $\E( \bm{z}^\intercal \bm{\theta} \mid x)$ via the least-squares regression:
\begin{equation}
    \label{eqn:isotonic-regression}
        \widehat{m} = \argmin_{m \in \mathcal F} \ 
        \frac{1}{2n_1}\sum_{i=1}^{n_1}\big(\bm z_{1i}^\intercal\widehat{\bm\theta} - m(x_{1i})\big)^2, 
\end{equation}
where $\mathcal{F}$ is a class of functions, and \eqref{eqn:isotonic-regression} includes various nonparametric methods, such as spline regression \citep{wahba1990spline}, and gradient boosting regression \citep{friedman2001greedy}. Then, $\widehat\rho$ is estimated base on the uncorrelatedness between $\bm{z}^\intercal \bm{\theta}$ and $w$, that is,
\begin{equation}
    \label{eqn:ratio_est}
    \frac{1}{n_1}\sum_{i=1}^{n_1} (\bm z_{1i}^\intercal\widehat{\bm\theta} ) 
    \left( \bm z_{1i}^\intercal\widehat{\bm\theta} - \widehat{\rho} \widehat{m}(x_{1i})\right) = 0, 
    \qquad \widehat{\rho} = \frac{\widehat{\bm{\theta}}^\intercal \sum_{i=1}^{n_1} ( \bm{z}_{1i} \bm{z}_{1i}^\intercal ) \widehat{\bm{\theta}} }{ \widehat{\bm{\theta}}^\intercal \sum_{i=1}^{n_1} \widehat{m}(x_{1i}) \bm{z}_{1i} }.
\end{equation}



Finally, the AIR estimator is $\widehat{\phi} = \widehat\rho\widehat{m}$. It is worth noting that AIR allows the estimation of a non-invertible transformation $\phi$, 
this is in contrast to the existing literature on data transformation 
(see \citet{yeo2000new}),
where only invertible transformations are considered.
In Section \ref{sec:sim}, the numerical results demonstrate the advantages of our method in detecting a quadratic relationship. For interval estimation of $\beta\phi(\cdot)$, there is ample literature devoted to constructing nonparametric confidence bands; see \citet{hall2013simple} and references therein.

\subsection{{Robustness to misspecified nonlinearity}}
The proposed model \eqref{model:structural-equation} considerably relaxes the linearity assumption in 2SLS.
Nevertheless, it is possible that the nonlinear transformation $\phi(\cdot)$ in \eqref{model:structural-equation} could be misspecified in practice, especially when two structural equations do not share the same transformation for the exposure:
\begin{equation}\label{model:misspecification}
        \phi(x) = \bm z^\intercal \bm\theta + w, \quad y = \beta \psi(x) + \bm z^\intercal \bm\alpha + \varepsilon,
\end{equation}
where $\phi\neq \psi$ are two different nonlinear functions. In TWAS, it is generally impossible to consistently estimate $\psi$ from the summary statistics. Yet, testing in Section \ref{sec:est_beta} remains valid.

\begin{corollary}
  Assume the conditions in Theorem \ref{theorem:2sir}, then 
  under $H_0$, \eqref{eqn:test-size} still holds for the model \eqref{model:misspecification}.
\end{corollary}


As a result, in our TWAS analysis, the p-values of the putative causal genes produced by 2SIR remain reliable regardless of whether the transformations are correctly specified. 
The simulation indicates that the proposed test enables control of the Type I error and outperforms its competitors in power in the misspecified cases; see Example 6 in Appendix \ref{sec:sim_misspecified}.

\section{Simulations}
\label{sec:sim}
This section examines the performance of the proposed 2SIR and AIR methods. {Moreover, for hypothesis testing, we propose to combine tests based on different slices, denoted as {Comb-2SIR}. Let $\mathcal{S}$ be a collection of candidate slices, we combine $p$-values based on different slices $S \in \mathcal{S}$ using the Cauchy combining method \citep{liu2020cauchy}.
More discussion about the Cauchy combining version of 2SIR over the number of slices is included in Appendix \ref{sec:comb}.} Specifically, the results are compared against 2SLS and PT-2SLS. For PT-2SLS, the optimal parameter $\lambda$ for minimizing skewness is estimated using maximum likelihood, c.f., Section 3 in \citet{yeo2000new}. 

The performance for both $\beta$ and $\phi(\cdot)$ are considered. Due to space constraints, this section only reports the performance of controlling Type I and II errors, coverage, and effectiveness of confidence intervals of $\beta$, details and results about $\phi(\cdot)$ estimation are provided in the Appendix \ref{sec:sim_link}.

\label{sec:sim_beta}

The simulated data $\mathcal{D} = (\bm{z}_i, x_i, y_i)_{i=1}^n$ is generated as follows. First, $\bm{z}_i$ is generated independently from $N(\bm{0}_p, \bm{\Sigma})$, and $w_i = u^2_i + \gamma_i$, where $u_i$ and $\gamma_i$ are independently generated from $N(0,1)$. 
Second, $x_i$ is generated as $x_i = \phi^{-1} ( \bm{\theta}^\intercal \bm{z}_i + w_i )$ when $\phi$ is invertible, and $x_i$ is randomly selected from the solution set $\{ x: \phi(x) = \bm{\theta}^\intercal \bm{z}_i + w_i \}$ when $\phi$ is non-invertible. 
Third, $y_i = \beta \phi(x_i) + \varepsilon_i$, where $\varepsilon_i = u_i + \zeta_i$, and $\zeta_i \sim N(0,1)$, thus $u_i$ acts as a confounder, and $w_i$ is dependent with $\varepsilon_i$. 
Finally, the first half of the data is provided as $\mathcal{D}_1$, and the summary data $\mathcal{D}_2$ is produced by the second half of the data to mimic the GWAS data. 
Six transformations are considered: (1) linear: $\phi(x) = x$; (2) logarithm: $\phi(x) = \log(x)$; (3) cube root: $\phi(x) = x^{1/3}$; (4) inverse: $\phi(x) = 1/x$; (5) piecewise linear: $\phi(x) = xI(x\leq 0) + 0.5 x I(x > 0)$; (6) quadratic: $\phi(x) = x^2$.

For Type I error and power analysis, we compute the proportions of rejecting out of 1,000 simulations under $H_0$ and out of 100 simulations under $H_a$, respectively. For constructing the CI, we report the averaged coverage and CI length out of 1,000 simulations. Note that the CIs for 2SLS and PT-2SLS are generated based on the asymptotic variance in \citet{inoue2010two}, the CIs for 2SIR are generated based on \eqref{eqn:confidence-interval}, and all CIs are left truncated at 0 since $\beta \geq 0$.

\begin{figure}[h]
    \centering
    \includegraphics[scale=.29]{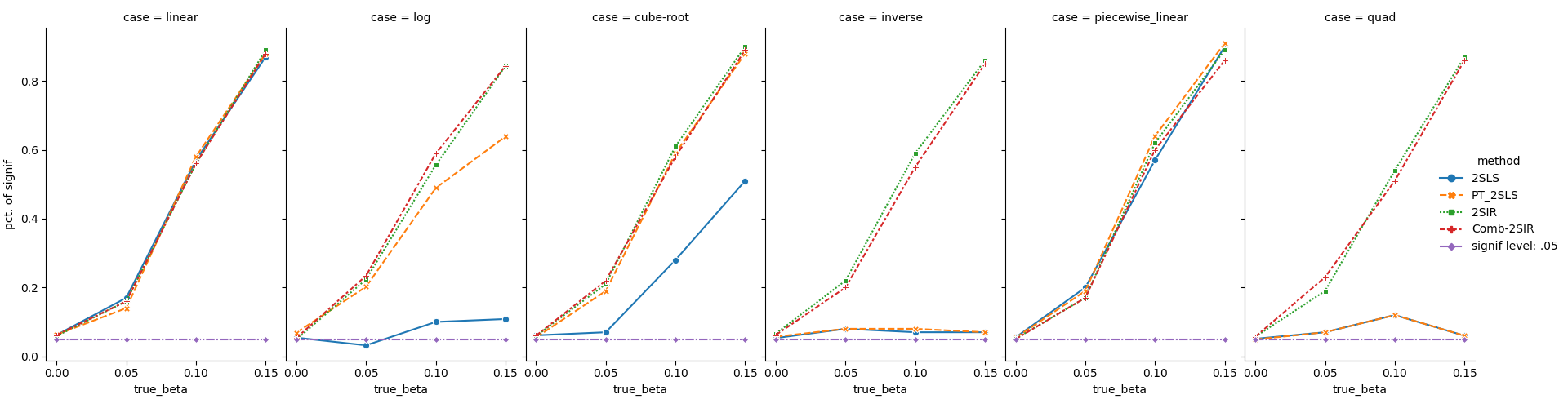}
    \includegraphics[scale=.29]{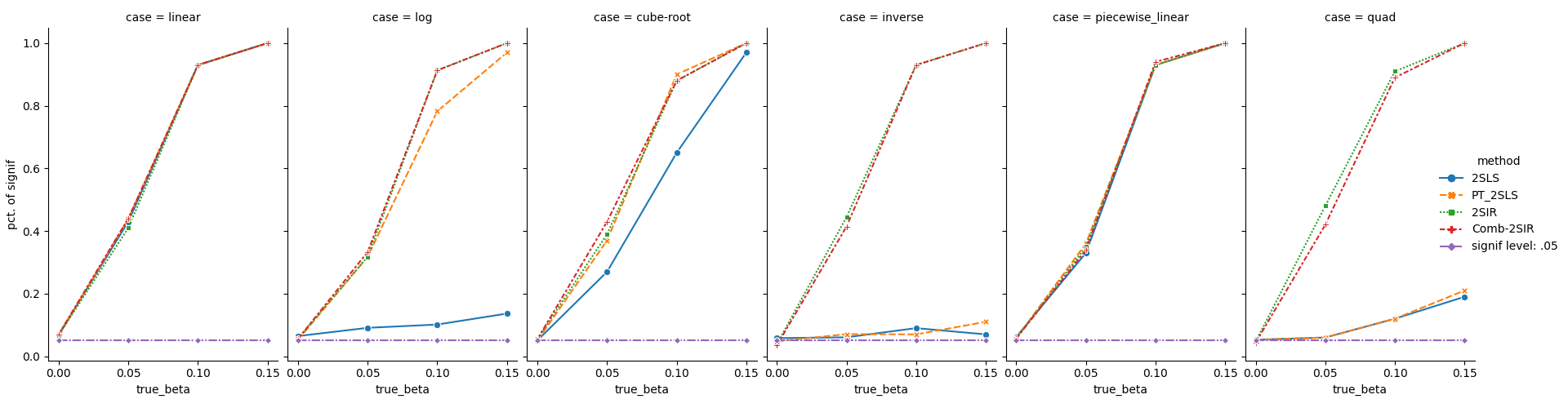}
    \includegraphics[scale=.29]{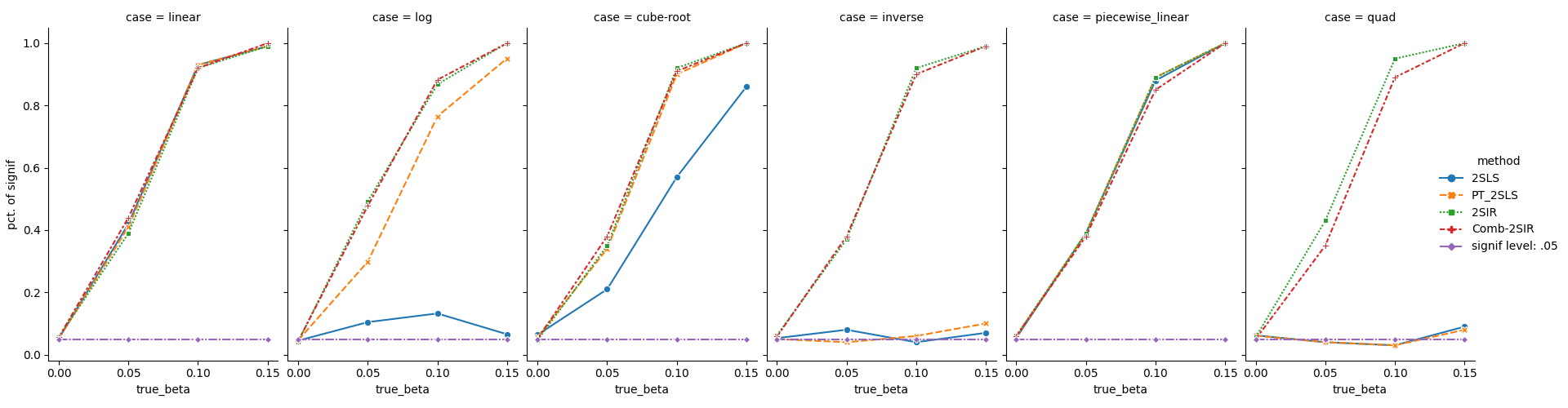}
    \caption{Empirical Type I error ($\beta_0 = 0$) and power ($\beta_0 > 0$) of marginal effect inference in Example 1 of Section \ref{sec:sim_beta}. $(n,p) = (2000, 50), (5000, 50), (5000, 100)$ from up to bottom. 
    }
    \label{fig:sim_beta}
\end{figure}

\noindent \textbf{Example 1} (Standard setting). In this example, we examine the proposed method under a standard setting. Specifically, we set $\bm{\Sigma} = \bm{I}_p$, $\bm{\theta} \sim N(\bm{0}, \bm{I}_p)$ and normalize it by its norm. 
We examine four cases: (i) $\beta = 0$, (ii) $\beta = .05$, (iii) $\beta = .10$, (iv) $\beta = .15$. Note that case (i) is for Type I error analysis, while $\beta > 0$ in (ii) - (iv), suggests power analysis. 
Moreover, the CI is produced based on (ii) $\beta = 0.05$. All empirical results are summarized in Figure \ref{fig:sim_beta} (testing) and Table \ref{tab:sim_CI} (CI).


\noindent \textbf{Examples 2-6}. Additional examples, including Example 2 (Invalid IVs), Example 3 (Categorical IVs), Example 4 (Weak IVs), Example 5 (Non-additive effects), Example 6 (Misspecified models) can be found in Appendix \ref{sec:implementation} to assess the performance of our methods under various data situations.

In summary, the simulation suggests the efficacy of the proposed 2SIR in managing all types of nonlinear transformations across various scenarios. The key conclusions are itemized below.
\begin{itemize}
    \setlength\itemsep{0em}
    \item For testing, as suggested in Figure \ref{fig:sim_beta}, the proposed 2SIR and its combined test yield competitive performance for ``linear'', ``cube-root'' and ``PL'' cases compared with 2SLS and PT-2SLS; and superior performance for ``log'', ``inverse'', and ``quad'' cases. 
    \item For CI, as indicated in Table \ref{tab:sim_CI}, 2SLS and PT-2SLS fail to provide valid CIs when ``inverse'' and ``quad'' transformations are used. For other cases, the proposed 2SIR yields competitive performance. In general, 2SIR is the only one that can provide a valid CI under an unknown nonlinear transformation.
    \item As suggested in Figures \ref{fig:sim_IV} - \ref{fig:sim_DE}, and Tables \ref{tab:sim_IV_CI} - \ref{tab:sim_CI_DE}, the proposed 2SIR continues to perform well with invalid, weak or categorical IVs. As indicated in Figure A.4 and Table A.4, 2SIR 
is also the most robust method against dominance and epistatic effects. As indicated in Figure \ref{fig:sim_mm}, the proposed methods can control Type I errors and are more powerful than the competitors when the transformation is misspecified.
\end{itemize}

\begin{table*}[!ht]
       \renewcommand{\arraystretch}{0.7}
       \centering
       \scalebox{0.75}{
       \begin{tabular}{@{}ccccccccccccccccccc@{}}
       \toprule
       & ~ & ~ & \multicolumn{2}{c}{\textit{2SLS}}  & ~ & ~ & \multicolumn{2}{c}{\textit{PT-2SLS}} & ~ & ~ & \multicolumn{2}{c}{\textit{2SIR} (proposed)} \\
       & $(n, p)$ & & coverage & length & && coverage & length &&& coverage & length \\
       \midrule
       & $(2000, 10)$ & linear & 0.944 & 0.132 &&& 0.943 & 0.132 &&& 0.967 & 0.138 \\
       & & log & 0.946 & 156.422 &&& 0.946 & 0.133 &&& 0.925 & 0.136 \\
       & & cube-root & 1.000 & 0.390 &&& 1.000 & 0.436 &&& 0.975 & 0.138 \\
       & & inverse & 0.964 & 0.522 &&& 0.930 & 0.134 &&& 0.979 & 0.138 \\
       & & PL & 0.950 & 0.134 &&& 0.949 & 0.134 &&& 0.971 & 0.138 \\
       & & quad & 0.831 & 0.093 &&& 0.823 & 0.092 &&& 0.951 & 0.139 \\
       \midrule
       & $(2000, 50)$ & linear & 0.941 & 0.128 &&& 0.943 & 0.129 &&& 0.974 & 0.136 \\
       & & log & 1.000 & 176.916 &&& 0.913 & 0.123 &&& 0.935 & 0.136 \\
       & & cube-root & 1.000 & 0.328 &&& 0.940 & 0.132 &&& 0.976 & 0.136 \\
       & & inverse & 0.990 & 0.149 &&& 0.882 & 0.096 &&& 0.979 & 0.131 \\
       & & PL & 0.943 & 0.126 &&& 0.944 & 0.127 &&& 0.982 & 0.134 \\
       & & quad & 0.743 & 0.084 &&& 0.743 & 0.083 &&& 0.976 & 0.134 \\
       \midrule
       & $(5000, 50)$ & linear & 0.950 & 0.094 &&& 0.952 & 0.095 &&& 0.978 & 0.095 \\
       & & log & 1.000 & 95.559 &&& 1.000 & 0.090 &&& 0.972 & 0.097 \\
       & & cube-root & 1.000 & 0.215 &&& 0.999 & 0.095 &&& 0.982 & 0.097 \\
       & & inverse & 0.801 & 0.209 &&& 0.640 & 0.060 &&& 0.972 & 0.096 \\
       & & PL & 0.951 & 0.096 &&& 0.960 & 0.096 &&& 0.977 & 0.096 \\
       & & quad & 0.522 & 0.052 &&& 0.523 & 0.051 &&& 0.976 & 0.095 \\
       \bottomrule
       \end{tabular}}
      \caption{Empirical coverage and length of the CI for in Example 1 of Section \ref{sec:sim_beta}.}
       \label{tab:sim_CI}
   \end{table*}

  \section{Real data analysis}
  In this section, we implement the proposed method for an analysis of the AD Neuroimaging Initiative (ADNI) dataset and the International Genomics of Alzheimer's Project (IGAP; \citet{lambert2013igap}) GWAS summary dataset to identify putative causal AD genes. 
  Specifically, the ADNI dataset consists of 819 individual-level subjects,
  17,201 genes, and 620,901 SNPs. The IGAP dataset consists of summary statistics of about 7 million SNPs to AD based on 54,162 samples.

  \label{sec:app}
  \begin{figure}[h]
      \centering
      \includegraphics[scale=.21]{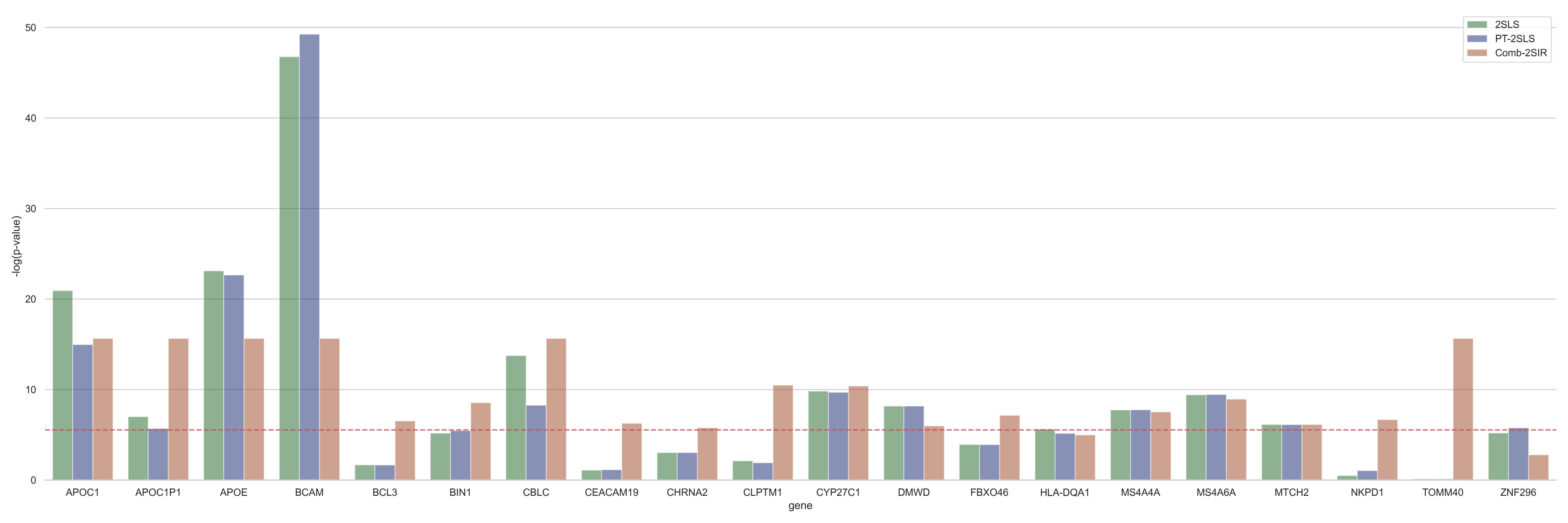}
      \includegraphics[width=0.46\textwidth]{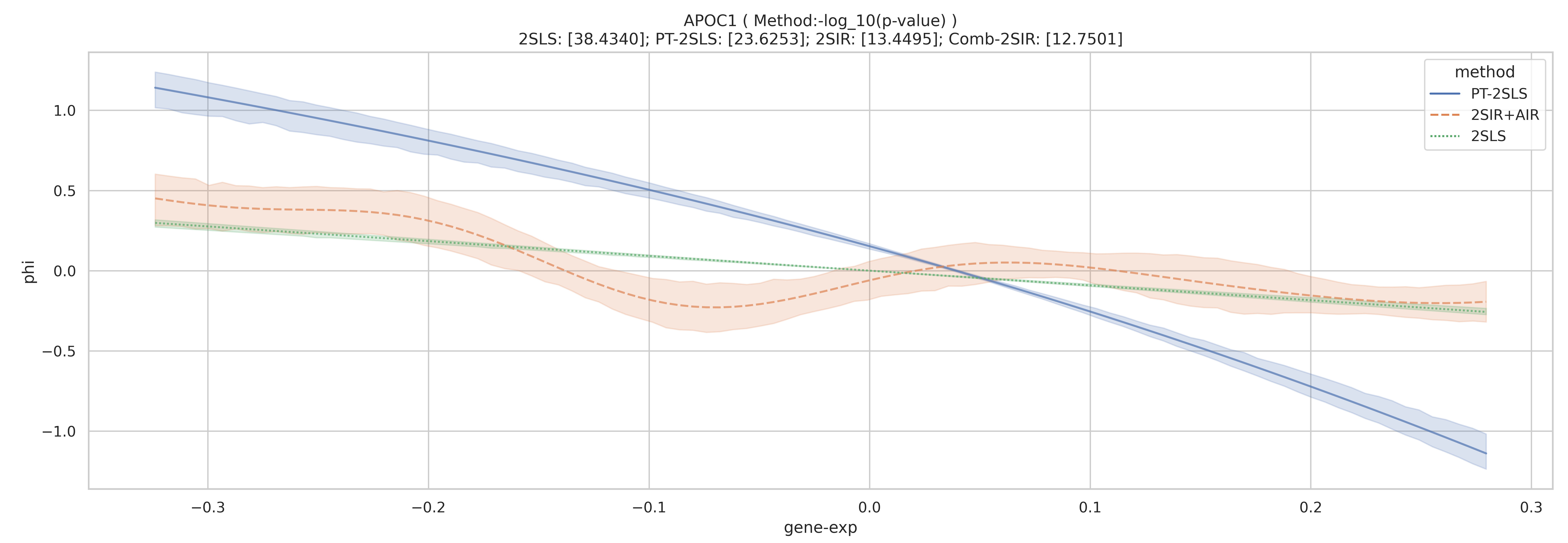}
      \includegraphics[width=0.46\textwidth]{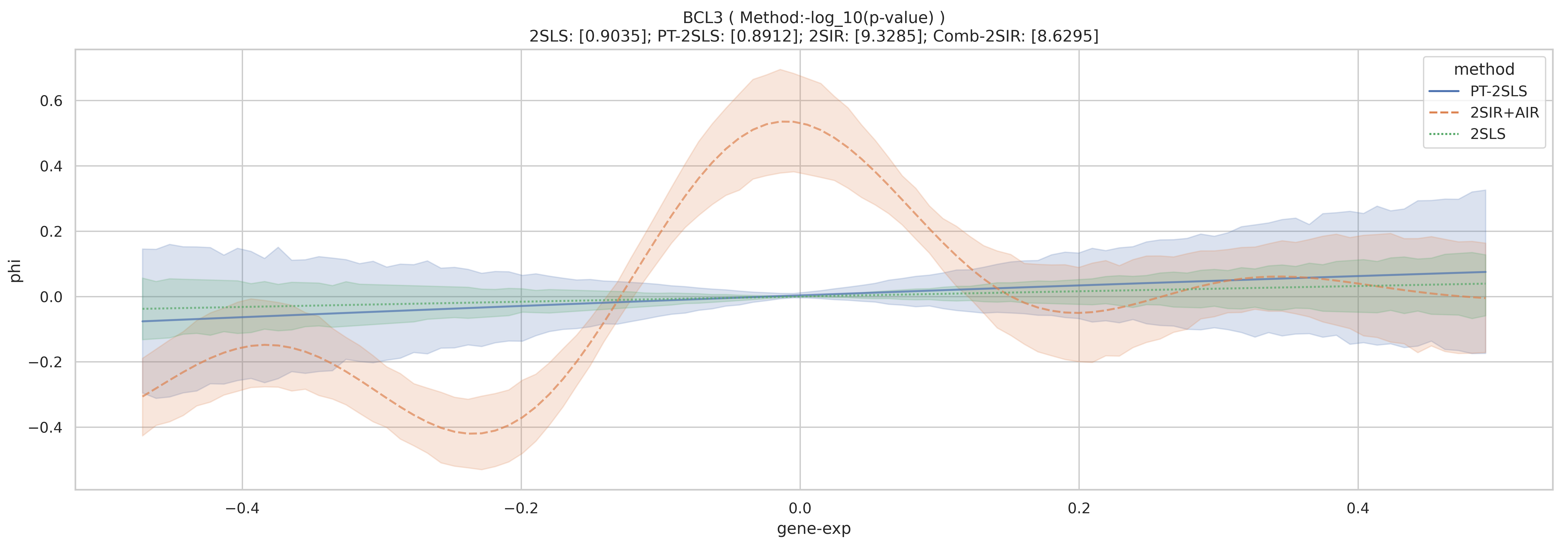}
      \includegraphics[scale=0.3]{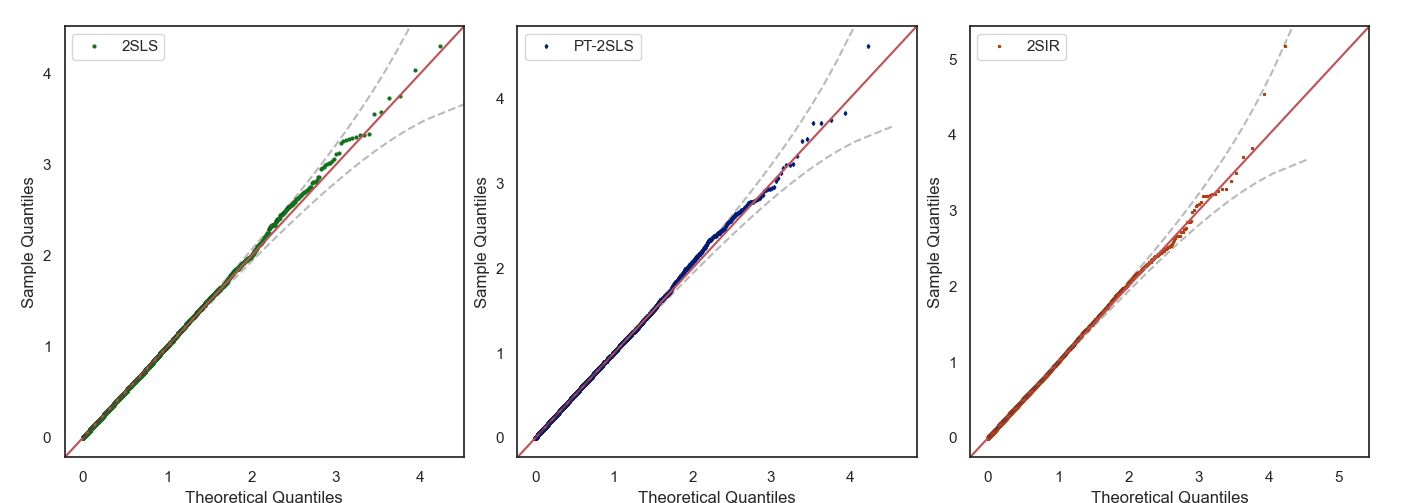}
      \caption{\textbf{Upper.} The bar-plot for significant AD genes, where the $x$-axis represents genes, the $y$-axis represents $-\log_{10}(p)$. \textbf{Middle.} Fitted transformations of two illustrative genes. APOC1 (\textit{left}) is identified by all methods. BCL3 (\textit{right}) is only identified by our method. \textbf{Lower.} QQ-plots for 2SLS, PT-2SLS, and 2SIR on ADNI negative control outcomes.}
      \label{fig:app_test}
      \vspace*{-.4cm}
  \end{figure}
  
  
  \noindent \textbf{Data preprocessing.} To facilitate the analysis, we pre-process the dataset and refine the candidate SNPs as follows. 
  For the ADNI dataset, we first exclude SNPs with MAF $\leq$ 0.05, with missing values, or failing the Hardy-Weinberg equilibrium test at the significant level of 0.001. Next, we further prune the SNPs to ensure that any of their pairwise Pearson correlations in absolute values were no more than 0.6. For the IGAP GWAS dataset, we conduct imputation for missing SNPs by using the software \texttt{ImpG} \citep{Pasaniuc2014ImpG}, based on 489 unrelated individuals with European ancestry from the 1000 Genomes Project \citep{1000genome}, yet remove the imputed SNPs with imputation accuracy smaller than 0.3. Finally, we define the cis-region of the gene by expanding 100kb upstream and downstream of its coding region, and take the top 50 intersecting SNPs (available both on the ADNI dataset and imputed IGAP dataset), with the largest absolute correlations with the gene's expression level. Taken together, the pre-processed dataset consists of 712 individual-level genotypes and gene expression with 50 SNPs and independent summary statistics for the associated SNPs based on 54,162 samples. 

  \noindent \textbf{Results.} Next, all methods are applied to the pre-processed data. 
  As indicated in Figure \ref{fig:app_test}, with the Bonferroni adjusted significance cutoff $0.05/17201$, \textcolor{black}{20} genes are identified as significantly related to AD by at least one method. 
  Specifically, among them \textcolor{black}{12} were significant by 2SLS and/or PT-2SLS, \textcolor{black}{18} are significant by Comb-2SIR.
  Two genes, APOE and TOMM40 on chromosome 19, are well-known to be related to AD \citep{bu2009apolipoprotein, mise2017tomm40, lyall2014alzheimer}; the former is identified by all \textcolor{black}{three} methods while the latter is only identified by Comb-2SIR.
  Besides TOMM40, 7 genes, BCL3, \textcolor{black}{BIN1}, CEACAM19, CHRNA2, CLPTM1, \textcolor{black}{FBXO46}, NKPD1, are only identified by Comb-2SIR. We searched these 7 genes in large-scale GWAS results and found all of them \textcolor{black}{except FBXO46} contained genetic variants that have been reported to be significantly associated with AD \citep{jansen2019genome, marioni2018gwas, beecham2014genome}.
  A further literature search gives more supporting evidence about their associations with AD. Specifically, BCL3 has been discovered to be associated with late-onset familial AD \citep{nho2017association, pericak1991linkage}; 
  \textcolor{black}{in AD brains, BIN1 has increased expression levels \citep{de2014alzheimer,chapuis2013increased};}
  CEACAM19 has been suggested as a candidate gene related to human aging \citep{evans2019identification};
  CHRNA2 has been implicated in potentially contributing to learning and memory functions \citep{nichol2015optogenetic} and as a potential target of clinical AD drugs \citep{cummings2019alzheimer}. 
  
  
  For illustration, Figure \ref{fig:app_test} (middle panel) shows the fitted transformations for two genes: APOC1 and BCL3 (others are included in Supplementary). For APOC1, which is successfully detected by 2SLS, the estimated 
  transformation by our method is roughly in agreement with the linear pattern estimated by 2SLS.  
  For BCL3, in contrast, the estimated transformation by our method is largely different from that of 2SLS and PT-2SLS, indicating that the linear pattern might be invalid here. This may be a reason for less significance given by {2SLS} and {PT-2SLS}, offering practical and empirical evidence for nonlinear causal effects in a real dataset.

  
  \noindent \textbf{Negative control outcomes.} {We also demonstrate Type I error control based on the ADNI dataset with negative control outcomes. Specifically, we implement the methods based on individual-level SNPs and gene expressions while generating negative control outcomes by simulating random noises so that \textit{no gene is causal to the outcome}. 
  In this case, the p-value is expected to follow a uniform distribution. 
  Figure \ref{fig:app_test} exhibits the QQ plots of the methods, suggesting that the p-values provided by 2SLS, PT-2SLS, and 2SIR are appropriately distributed in this negative control dataset. }
  
  \section{Discussion and conclusions}
  
  Nonlinear modeling in TWAS has potential significance in identifying causal gene-trait associations. However, it {is plagued by 
  the lack of individual-level GWAS data} (with only summary statistics for the outcome available). 
  In this paper, we have proposed a flexible causal model for summary data while allowing an arbitrary nonlinear causal effect, substantially relaxing the assumption of linearity in the current practice of TWAS.
  A novel method called 2SIR+AIR is developed to estimate the marginal causal effect and the nonlinear transformation, covering 2SLS as a special case. 
  In addition, we have developed inferential tools to assess exposure-outcome associations, including hypothesis testing and interval estimation; in particular, our test is robust to model misspecification. 
  
  We have demonstrated the applicability of the proposed model and methods by studying the ADNI gene expression and the IGAP GWAS datasets to identify putative causal genes for AD. Our results suggest that the proposed method agrees with two existing methods (2SLS and PT-2SLS) in 10 of 12 putative causal genes, but it additionally identifies 7 other potential AD genes. 
  We also observe higher $R^2$'s for the stage one model of our method than existing models, offering another source of evidence that nonlinear causal effects are likely to be present in real data. 
  Our finding reasonably suggests potential nonlinearity in gene-trait causal associations based on GWAS data. 
  We believe that the proposed method has great potential and could further advance research in TWAS, including nonlinear treatment effect analysis, subgroup analysis, and robustness analysis.
  Finally, in addition to TWAS, the proposed method can be equally applied to study other exposure-outcome causal relationships in a more general context.

\acks{We thank a bunch of people and funding agencies.}

\bibliography{ref}

\appendix

\numberwithin{equation}{section}
\numberwithin{table}{section}
\numberwithin{figure}{section}

\section{Simulation for transformation estimation}
\label{sec:sim_link}
This subsection examines the proposed adjusted inverse regression (\textit{2SIR+AIR}) in \eqref{eqn:phi_eva} under various nonlinear transformations, and the estimation accuracy is measured by mean square error (MSE) and uniform error (UE):
\begin{equation}
    \label{eqn:phi_eva}
    \text{MSE}(\widehat{\phi}, \phi_0) = \E\Big( \big( \widehat{\phi}(x) - \phi_0(x) \big)^2 \Big), \qquad \text{UE}(\widehat{\phi}, \phi_0) = \E\sup_{x \in \mathcal{X}} \big| \widehat{\phi}(x) - \phi_0(x) \big| 
\end{equation}
where $\mathcal{X}$ is a region of causal interest, which is replaced as 100 grid points of [5\%-quantile, 95\%-quantile] of $x$ for evaluation. We also compare the results with a conditional mean function to highlight the role of the ratio correction in \eqref{eqn:ratio_est}.

\begin{figure}
 \centering
 \includegraphics[width=0.45\textwidth]{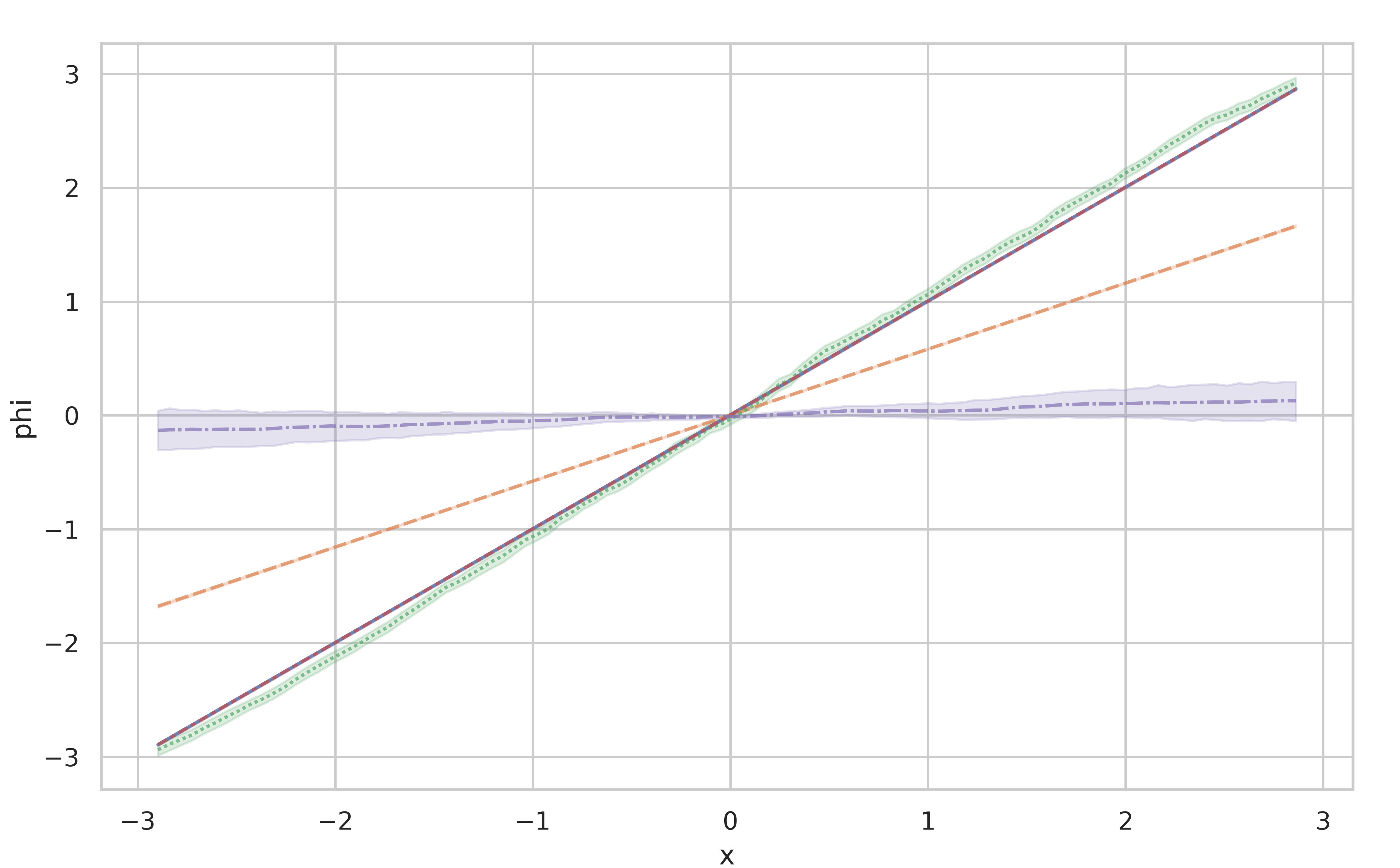}
 \includegraphics[width=0.45\textwidth]{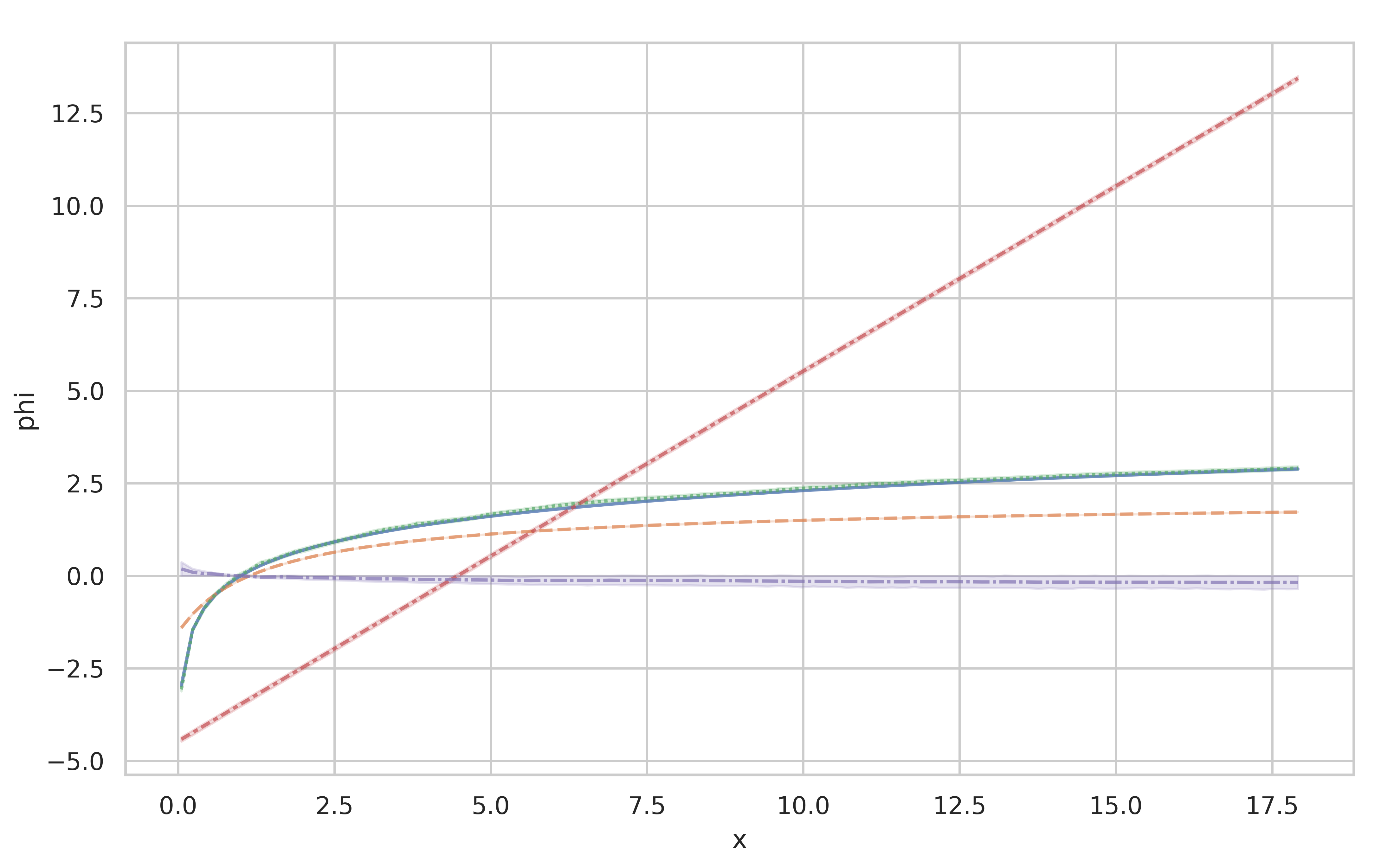}
 \includegraphics[width=0.45\textwidth]{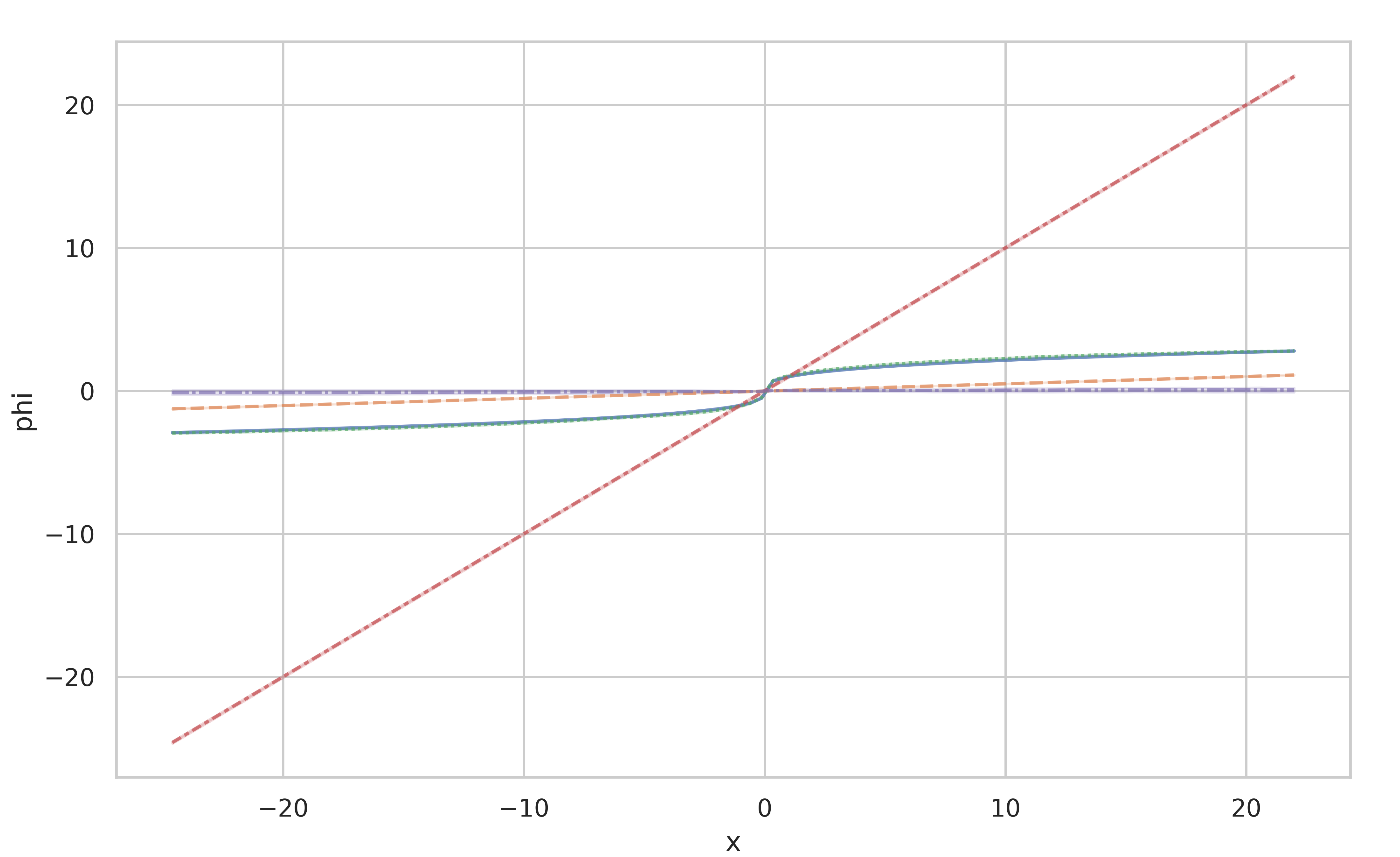}
 \includegraphics[width=0.45\textwidth]{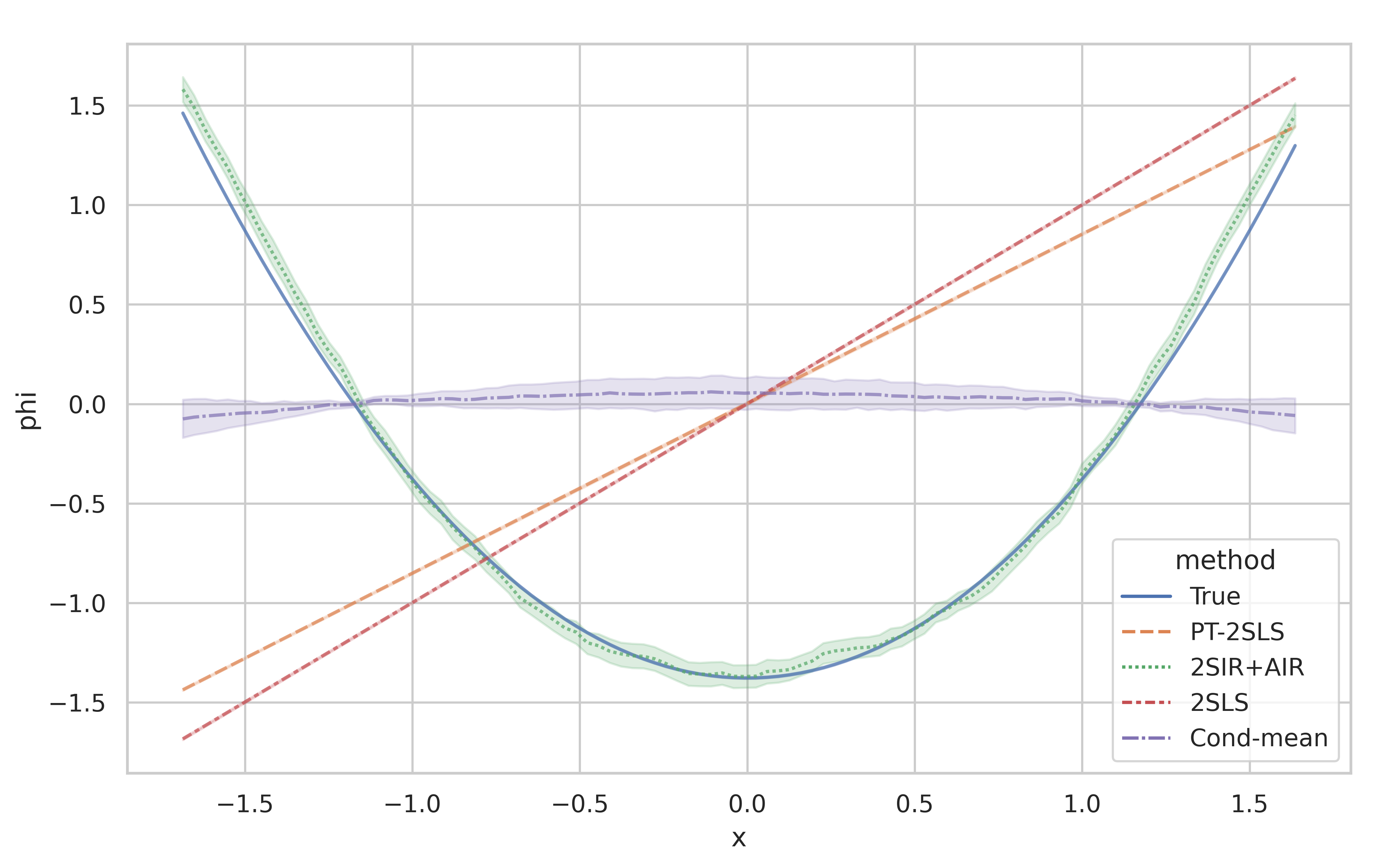}
 \caption{Fitted transformations of the simulated example in Section \ref{sec:sim_link}, where the true transformations are: (1,1) linear; (1,2) logarithm; (2,1) cubic root; (2,2) quadratic.}
 \label{fig:sim_link}
\end{figure}

Specifically, we set $\bm{\theta} = ( p^{-1/2}, \cdots, p^{-1/2} )^\intercal$ and $\beta = 1$ in \eqref{model:structural-equation}.
Note that $\mathcal{D}_1$ and $\mathcal{D}_2$ are generated with the same setting in Example 1 in Section \ref{sec:sim_beta} with $w_i = u_i + \gamma_i$, $u_i$ and $\gamma_i$ are independently generated from $N(0,1)$. 
Five nonlinear transformations are considered: (1) linear: $\phi(x) = x$; (2) logarithm: $\phi(x) = \log(x)$; (3) cube root: $\phi(x) = x^{1/3}$, (4) piecewise linear (PL): $\phi(x) = x I(x\leq 0) + 0.5x I(x > 0)$, (5) quadratic (quad): $\phi(x) = x^2$. Note that the conditional mean regression \eqref{eqn:isotonic-regression} is conducted based on a KNN model with the number of neighbors as 100.
The simulation is replicated 100 times with $n=2000, p=10, 50, 100$, the resulting MSEs and UEs are summarized in Table \ref{tab:sim_link}, and the fitted transformations for $p=10$ is illustrated in Figure \ref{fig:sim_link}. 

It is evident that the proposed 2SIR+AIR method substantially outperforms 2SLS and PT-2SLS in most cases, except that 2SLS yields better performance in the ``linear'' case where the proposed model in \eqref{model:structural-equation} becomes a linear structural equation model. For other cases, the amount of improvement is significant, with the largest improvement of (MSE: 99.9\%, UE: 96.6\%) and (MSE: 92.2\%, UE: 64.4\%) over 2SLS and PT-2SLS, respectively.

\begin{table*}[h]
    \centering
    \scalebox{.8}{
    \begin{tabular}{@{}ccccccccccccccccccc@{}} 
    \toprule
    & ~ & \multicolumn{2}{c}{\textit{2SLS}}  & ~ & \multicolumn{2}{c}{\textit{PT-2SLS}} \\
    $p$ & & MSE & UE && MSE & UE
    \\
    \midrule
    $10$ & linear & 0.000(.000) & 0.000(.000) & ~ & 0.525(.005) & 1.216(.005) \\ 
    ~ & log & 363.405(48.756) & 9.892(0.045) & ~ & 0.619(.004) & 1.362(.004) \\ 
    ~ & cube-root & 346.575(6.023) & 21.777(0.042) & ~ & 1.293(.009) & 1.737(.009) \\ 
    ~ & PL & 1.026(.005) & 2.130(.002) & ~ & 0.540(.004) & 1.284(.005) \\ 
    ~ & quad & 2.461(.009) & 3.073(.004) & ~ & 2.083(.009) & 2.824(.004) \\ 
    $50$ & linear & 0.000(.000) & 0.000(.000) & ~ & 0.535(.005) & 1.171(.004) \\ 
    ~ & log & 223.565(22.881) & 12.106(.028) & ~ & 0.616(.004) & 1.342(.003) \\ 
    ~ & cube-root & 355.761(5.317) & 19.961(.038) & ~ & 1.302(.010) & 1.738(.008) \\ 
    ~ & PL & 1.022(.004) & 2.134(.002) & ~ & 0.546(.005) & 1.256(.005) \\ 
    ~ & quad & 2.474(.009) & 3.287(.003) & ~ & 2.095(.008) & 3.033(.004) \\ 
    $100$ & linear & 0.000(.000) & 0.000(.000) & ~ & 0.526(.004) & 1.204(.004) \\ 
    ~ & log & 615.467(32.895) & 7.429(.044) & ~ & 0.623(.005) & 1.580(.004) \\ 
    ~ & cube-root & 354.663(5.198) & 20.571(.023) & ~ & 1.300(.009) & 1.740(.010) \\ 
    ~ & PL & 1.018(.005) & 2.103(.002) & ~ & 0.541(.004) & 1.176(.004) \\ 
    ~ & quad & 2.468(.009) & 3.097(.004) & ~ & 2.092(.008) & 2.851(.004) \\
    \midrule 
    ~ & ~ & ~ & ~ & ~ & ~ & ~ \\
    \toprule
    & ~ & \multicolumn{2}{c}{\textit{Cond-mean(KNN)}} & ~ & \multicolumn{2}{c}{\textit{2SIR+AIR} (proposed)} \\
    $p$ & & MSE & UE && MSE & UE\\
    \midrule
    $10$ & linear & 3.530(.179) & 3.076(.090) & ~ & 0.117(.003) & 0.615(.012) \\ 
    ~ & log & 3.471(.205) & 2.945(.094) & ~ & 0.118(.002) & 0.589(.016) \\ 
    ~ & cube-root & 3.336(.205) & 2.766(.099) & ~ & 0.113(.002) & 0.584(.016) \\ 
    ~ & PL & 2.853(.207) & 2.614(.104) & ~ & 0.123(.003) & 0.645(.016) \\ 
    ~ & quad & 1.323(.060) & 1.568(.042) & ~ & 0.123(.004) & 0.638(.013) \\ 
    $50$ & linear & 3.305(.214) & 3.022(.096) & ~ & 0.125(.003) & 0.598(.015) \\ 
    ~ & log & 3.273(.214) & 2.829(.104) & ~ & 0.124(.002) & 0.534(.016) \\ 
    ~ & cube-root & 3.408(.216) & 2.922(.100) & ~ & 0.121(.003) & 0.561(.013) \\ 
    ~ & PL & 3.113(.214) & 2.965(.100) & ~ & 0.119(.003) & 0.583(.016) \\ 
    ~ & quad & 1.162(.069) & 1.581(.055) & ~ & 0.163(.006) & 0.837(.020) \\ 
    $100$ & linear & 3.203(.217) & 3,019(.095) & ~ & 0.142(.003) & 0.570(.010) \\ 
    ~ & log & 3.591(.220) & 2.741(.111) & ~ & 0.148(.003) & 0.539(.011) \\ 
    ~ & cube-root & 3.818(.217) & 3.157(.104) & ~ & 0.140(.003) & 0.565(.012) \\ 
    ~ & PL & 3.638(.219) & 3.057(.107) & ~ & 0.142(.003) & 0.572(.015) \\ 
    ~ & quad & 1.201(.076) & 1.492(.057) & ~ & 0.232(.009) & 1.015(.023) \\ 
    \bottomrule
    \end{tabular}}
    \caption{Mean square error (MSE) and uniform error (UE) (standard errors in parentheses) for the simulated example in Section \ref{sec:sim_link}. Here cond-mean({KNN}), and {2SIR+AIR} denote nonparametric regression in \eqref{eqn:isotonic-regression}, and the proposed method in \eqref{eqn:ratio_est}, respectively. }
    \label{tab:sim_link}
\end{table*}

\section{Implementation and additional simulations}\label{sec:implementation} 
\subsection{Computation and hyperparameter tuning}\label{sec:computation}

To solve (4), we first approximate the $\|\cdot\|_0$ penalty by the SCAD \citep{fan2001variable}, and then consider the corresponding regularized problem:
\begin{equation}\label{eqn:scad-regression}
    \begin{split}
        \min_{\bm\alpha,\beta} \ 
        (\widehat{\bm\theta}\beta + \bm\alpha)^\intercal\bm Z_2^\intercal \bm Z_2 (\widehat{\bm\theta}\beta + \bm\alpha) 
        - 2\bm Y_2^\intercal\bm Z_2 (\widehat{\bm\theta}\beta + \bm\alpha)
        + \lambda p_{a}(\bm\alpha),
    \end{split}
\end{equation}
where $p_a(\bm\alpha) = \sum_{j=1}^p p_a(\alpha_j)$ is the SCAD penalty, 
$\lambda > 0$ is a tuning parameter controlling the sparsity of the solution, 
and $a>0$ is a parameter in the SCAD, c.f. \eqref{eqn:SCAD}. 
For each choice of $\lambda$, the solution $(\widehat{\bm\alpha}_{\lambda},\widehat{\beta}_{\lambda})$ of
\eqref{eqn:scad-regression} can be efficiently computed by the local linear approximation algorithm \citep{zou2008one}.
{Next, fixing $K$, we refit an ordinary least squares (OLS) regression with $\widehat{\bm\theta}^\intercal \bm z$ and the top $K$ variables in $\widehat{\bm\alpha}_{\lambda}$ for each $\lambda$. Let $(\widehat{\bm\alpha}_{\lambda,K},\widehat{\beta}_{\lambda,K})$ be the resulting OLS estimate. Then define 
\begin{equation*}
    (\widehat{\bm\alpha}_{K},\widehat{\beta}_{K}) = \argmin_{(\widehat{\bm{\alpha}}_{\lambda,K}, \widehat{\beta}_{\lambda,K})} \text{RSS}_2(\widehat{\bm{\alpha}}_{\lambda,K}, \widehat{\beta}_{\lambda,K})
\end{equation*} 
as the solution to (4), where $\text{RSS}_2(\widehat{\bm{\alpha}}_{\lambda,K}, \widehat{\beta}_{\lambda,K})$ is the residual sum of squares.

To choose the best performing $K$, we use BIC for tuning criteria. Specifically, define
\begin{align*}
    \widehat{\text{BIC}}(K) &= \frac{ \text{RSS}_2(\widehat{\bm{\alpha}}_{K}, \widehat{\beta}_{K}) }{\widehat{\sigma}^2_e} + {\log(n_2)} (K+1),
\end{align*}
where $\widehat{\sigma}^2_e = \text{RSS}_2( \widehat{\bm \alpha}_\text{ols}, 0 ) / n_2$ is an estimate of $\sigma^2_e$ in \eqref{model:y-z} and $\widehat{\bm\alpha}_{\text{ols}} = (\bm Z_2^\intercal\bm Z_2)^{-1}\bm Z_2^\intercal\bm Y_2$. Then we choose $K$ that minimizes $\widehat{\text{BIC}}(K)$, and use $(\widehat{\bm\alpha}_{K},\widehat{\beta}_K)$ for the subsequent data analysis.}


\subsection{Stability combination of p-values}
\label{sec:comb}

\begin{figure}
        \centering
        \includegraphics[scale=.065]{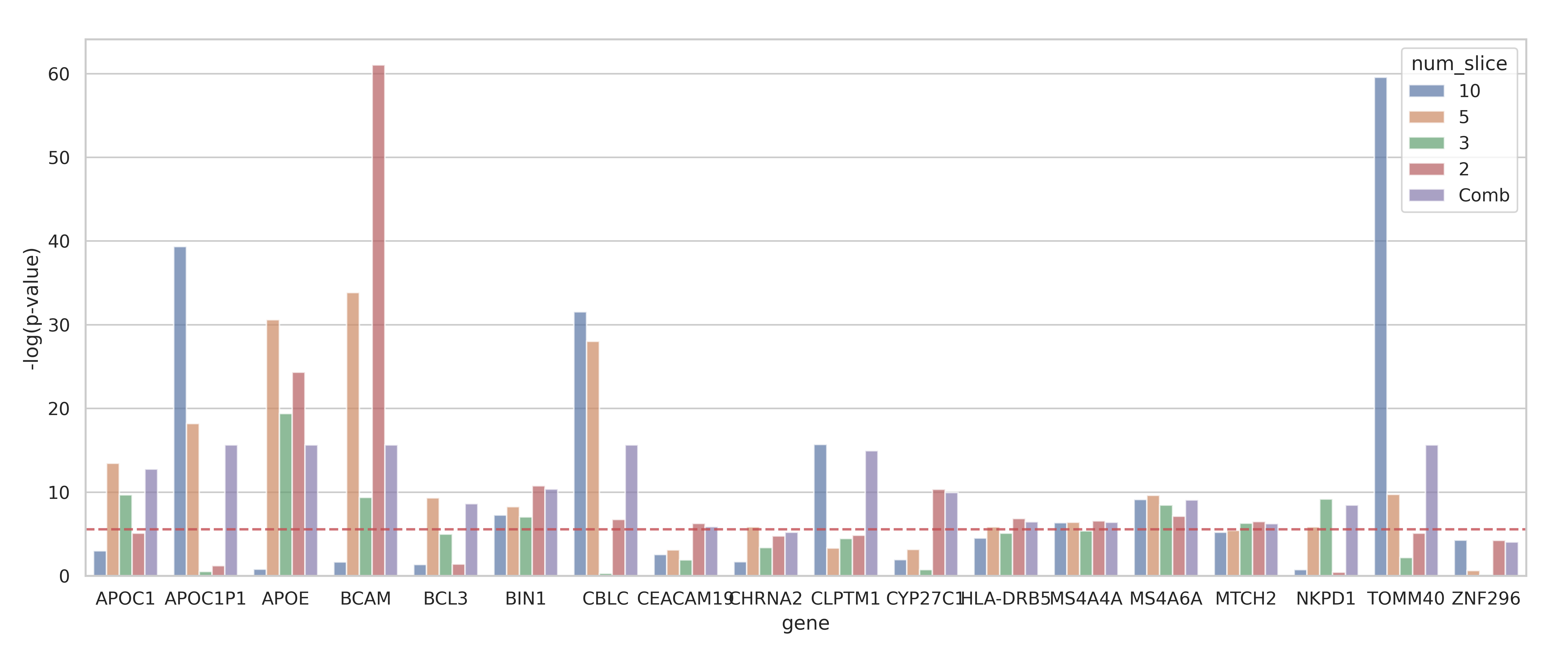}
        \caption{The bar-plot for the negative logarithm of $p$-values of significant genes in Section 4 with different numbers of slices ($S=2, 3, 5, 10$) based on the proposed method.}
        \label{fig:app_comb}
    \end{figure}

In (8), the slicing scheme is treated as fixed.
Although the number of slices $S$ has been regarded as a hyperparameter of minor importance \citep{li1991sliced,cook2009regression}, our experiments and existing literature \citep{becker2007note} suggest that the numerical results may vary greatly as $S$ changes. Specifically, we produce $p$-values for significant genes in Section 4 with a different number of slices based on the proposed method. Figure \ref{fig:app_comb} clearly suggests that $p$-values significantly affected by the number of slices ($S = 2, 3, 5, 10$). 
Hence, a gap in the choice of $S$ exists between theory and practice. 

To bridge this gap, we propose to combine the tests based on different slicing schemes.
Specifically, let $\mathcal S$ be a collection of candidate slicing schemes. We combine $p$-values based on different slices $S \in \mathcal{S}$ based on the  Cauchy combination method \citep{liu2020cauchy}:
\begin{equation}\label{eqn:combine}
        p_* = 0.5 - (\arctan t_0) / \pi, \quad  t_0 = \sum_{S \in \mathcal{S}} w_i \tan\Big( \big(0.5 - P_W( |W| > \widehat T_S ) \big) \pi \Big),
\end{equation}
where the weights $w_i$s are nonnegative and $\sum_{i=1}^{|\mathcal{S}|} w_i = 1$, $\widehat T_S$ is the test statistic in (8) with the subscript emphasizing its dependence on $S$,
and $W\sim N(0,1)$ is a standard normal variable independent of the data.
For illustration, we focus on a combined version of the proposed method with $w_i = 1/|\mathcal{S}|$.  
Note that we could apply other types of combining such as order statistics of the $p$-values, and corrected arithmetic and geometric means \citep{vovk2020combining}.

\subsection{Simulation results for Invalid IVs with or without correlated pleiotropy}
\begin{figure}[h]
  \centering
  \includegraphics[scale=.3]{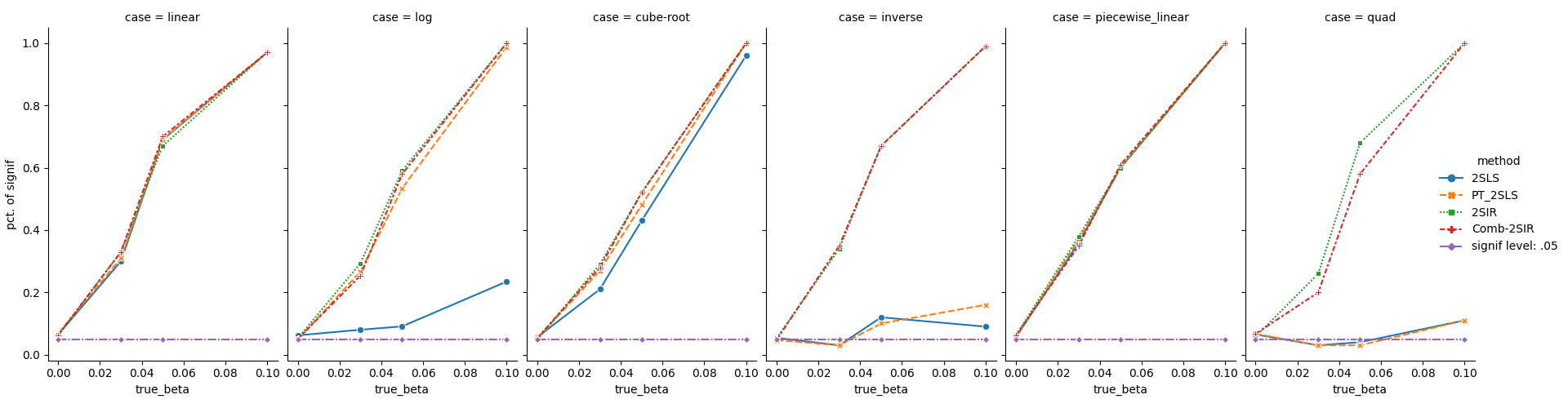}
  \includegraphics[scale=.3]{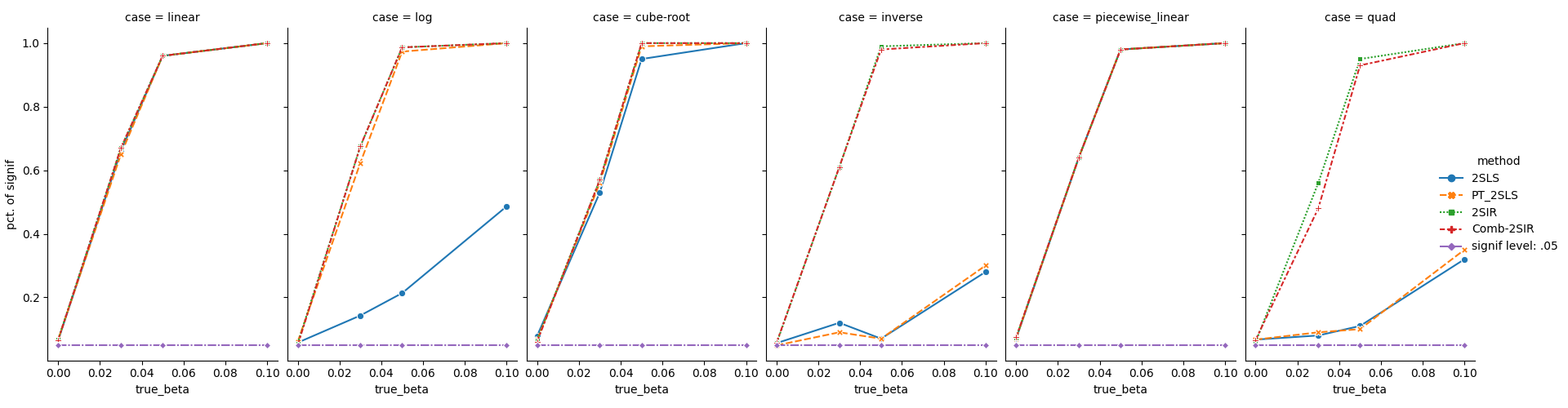}
  \caption{Empirical Type I error ($\beta_0 = 0$) and power ($\beta = 0.05, 0.10, 0.15$) for the simulated example (invalid IVs) in Example 2. $\nu = 0.0, 0.5$ from top to bottom.}
  \label{fig:sim_IV}
\end{figure}

\begin{figure}[h!]
    \centering
    \includegraphics[scale=.3]{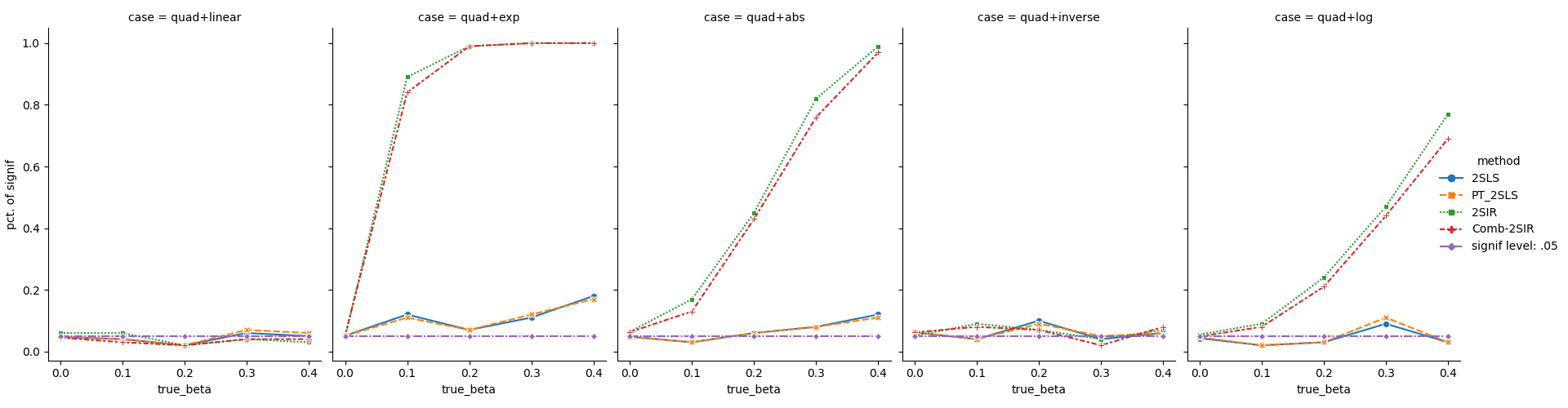}
    \caption{Empirical Type I error ($\beta_0 = 0$) and power ($\beta = 0.05, 0.10, 0.15$) of the proposed nonlinear causal test for the simulated example (invalid IVs with correlated pleiotropy) in Example 2 of Section \ref{sec:sim}.}
    \label{fig:sim_cor_pleiotropy}
\end{figure}

\noindent \textbf{Example 2} (Invalid IVs). In this example, we examine the proposed method with invalid IVs. 
Specifically, $\bm{z}_i$ is generated with $\Sigma_{ij} = \nu^{|i-j|}$. Then, $x_i$ is generated based on the same procedure in Example 1. Finally, $y_i$ is generated as $y_i = \beta \phi(x_i) + \bm{\alpha}^\intercal \bm{z}_i + \epsilon_i$. Here $\bm{\alpha} = (1, 1, 1, 1, 1, 0, \cdots, 0)$ indicates that the first five elements are invalid IVs. We examine four cases: (i) $\beta = 0$, (ii) $\beta = .03$, (iii) $\beta = .05$, (iv) $\beta = .10$. We construct CIs for (iii) $\beta = .05$.
All empirical results are summarized in Figure \ref{fig:sim_IV} (testing) and Table \ref{tab:sim_IV_CI} (CI) based on $(n,p) = (10000, 50)$, and $\nu = 0.0, 0.5$. Moreover, we further consider invalid IVs with correlated pleiotropy, where $\bm{\theta} = \bm{\theta}_0 + \bm{\mu} $ and $\bm{\alpha} = \bm{\alpha}_0 + \bm{\mu}$ where $\bm{\theta}_0$ and $\bm{\alpha}_0$ are simulated with the same procedure in Example 1, and $\bm{\mu} = (\mu_1, \cdots, \mu_5, 0, \cdots)^\intercal$ with $\mu_j \sim N(0,1)$. All empirical results are summarized in Figure \ref{fig:sim_cor_pleiotropy} (testing) based on $n = 10000, p = 50$, and $\nu = 0.5$.

\begin{table*}[h]
  \centering
  \renewcommand{\arraystretch}{0.7}
  \scalebox{0.75}{
  \begin{tabular}{@{}ccccccccccccccccccc@{}} 
  \toprule
  & ~ & ~ & \multicolumn{2}{c}{\textit{2SLS}}  & ~ & ~ & \multicolumn{2}{c}{\textit{PT-2SLS}} & ~ & ~ & \multicolumn{2}{c}{\textit{2SIR} (proposed)} & \\
  & $\nu$ & & coverage & length & && coverage & length &&& coverage & length \\
  \midrule
  & $0.0$ & linear & 0.945 & 0.078 &&& 0.945 & 0.078 &&& 0.948 & 0.078 \\
  & & log & 0.999 & 79.988 &&& 0.928 & 0.078 &&& 0.952 & 0.078 \\
  & & cube-root & 0.965 & 0.190 &&& 0.972 & 0.079 &&& 0.949 & 0.079 \\
  & & inverse & 0.598 & 0.159 &&& 0.510 & 0.050 &&& 0.954 & 0.078  \\
  & & PL & 0.951 & 0.079 &&& 0.950 & 0.079 &&& 0.951 & 0.079 \\
  & & quad & 0.443 & 0.043 &&& 0.456 & 0.043 &&& 0.964 & 0.079 \\
  \midrule
  & $0.5$ & linear & 0.951 & 0.050 &&& 0.951 & 0.050 &&& 0.945 & 0.050 \\
  & & log & 1.000 & 213.678 &&& 0.948 & 0.056 &&& 0.946 & 0.050 \\
  & & cube-root & 1.000 & 0.216 &&& 0.945 & 0.051 &&& 0.943 & 0.049 \\
  & & inverse & 0.827 & 0.210 &&& 0.645 & 0.062 &&& 0.940 & 0.050 \\
  & & PL & 0.942 & 0.050 &&& 0.912 & 0.049 &&& 0.936 & 0.050  \\
  & & quad & 0.541 & 0.055 &&& 0.514 & 0.055 &&& 0.946 & 0.049 \\
  \bottomrule
  \end{tabular}}
  \caption{Empirical coverage and length of the CI for in Example 2 (invalid IVs). }
  \label{tab:sim_IV_CI}
\end{table*}


\subsection{Simulation results for categorical IVs}
\noindent \textbf{Example 3} (Categorical IVs). Note that the proposed method requires that the IVs follow an elliptically symmetric distribution, which is usually invalid for categorical data. Yet, in practice, a categorical IV is often involved in causal inference, such as SNP data. In this example, we examine if the proposed method can be applied to categorical IVs.
Specifically, the IVs $(\bm{z}_i)_{i=1, \cdots, n}$ are generated as $\bm{z}_i = \bm{\tau}_i + \bm{\tau}_i'$ to mimic the SNP data, where $\bm{\tau}_i$ and $\bm{\tau}_i'$ are independent Bernoulli trials, each with a probability of success $0.3$. Moreover, we set $\bm{\theta} \sim N(\bm{0}, \bm{I}_p)$ and normalize it by its norm, then $x_i$ and $y_i$ are generated following the same procedure in Example 1. All empirical results are summarized in Table \ref{tab:sim_cate} (testing), Table \ref{tab:sim_CI_cate} (CI), and Figure \ref{fig:sim_beta} (boxplot).

\begin{table*}[h]
    \centering
    \scalebox{.58}{
    \begin{tabular}{@{}ccccccccccccccccccc@{}} 
    \toprule
    & ~ & ~ & \multicolumn{2}{c}{\textit{2SLS}}  & ~ & ~ & \multicolumn{2}{c}{\textit{PT-2SLS}} & ~ & ~ & \multicolumn{2}{c}{\textit{2SIR} (proposed)} & ~ & ~ & \multicolumn{2}{c}{\textit{Comb-2SIR} (proposed)} \\
    & $(n, p)$ & & Type I & Power & && Type I & Power &&& Type I & Power &&& Type I & Power \\
    \midrule
    & $(2000,10)$ & linear & .040 & (0.20, 0.37, 0.47) &&& .040 & (0.20, 0.39, 0.51) &&& .048 & (0.20, 0.40, 0.52) &&& .046 & (0.18, 0.40, 0.54) \\
    & & log & .050 & (0.03, 0.14, 0.14) &&& .058 & (0.07, 0.25, 0.52) &&& .055 & (0.06, 0.23, 0.60) &&& .057 & (0.09, 0.22, 0.60) \\
    & & cube-root & .052 & (0.08, 0.16, 0.36) &&& .055 & (0.10, 0.32, 0.54) &&& .057 & (0.13, 0.35, 0.53) &&& .054 & (0.13, 0.35, 0.55) \\
    & & inverse & .050 & (0.05, 0.07, 0.15) &&& .044 & (0.03, 0.08, 0.12) &&& .060 & (0.14, 0.24, 0.57) &&& .066 & (0.14, 0.25, 0.58) \\
    & & PL & .058 & (0.13, 0.30, 0.52) &&& .059 & (0.15, 0.29, 0.52) &&& .055 & (0.15, 0.28, 0.51) &&& .055 & (0.14, 0.30, 0.52) \\
    & & quad & .051 & (0.08, 0.02, 0.14) &&& .053 & (0.08, 0.03, 0.15) &&& .040 & (0.10, 0.23, 0.58) &&& .043 & (0.10, 0.17, 0.56) \\
    \midrule
    & $(2000, 50)$ & linear & .070 & (0.11, 0.19, 0.52) &&& .069 & (0.13, 0.21, 0.56) &&& .060 & (0.10, 0.18, 0.59) &&& .062 & (0.11, 0.18, 0.59) \\
    & & log & .065 & (0.04, 0.08, 0.12) &&& .062 & (0.08, 0.15, 0.28) &&& .061 & (0.08, 0.18, 0.46) &&& .063 & (0.06, 9.18, 0.47)  \\
    & & cube-root & .059 & (0.05, 0.09, 0.14) &&& .061 & (0.04, 0.18, 0.46) &&& .042 & (0.07, 0.21, 0.54) &&& .045 & (0.06, 0.24, 0.49) \\
    & & inverse & .050 & (0.05, 0.06, 0.06) &&& .055 & (0.04, 0.11, 0.09) &&& .059 & (0.08, 0.25, 0.49) &&& .069 & (0.10, 0.21, 0.45) \\
    & & PL & .050 & (0.09, 0.26, 0.50) &&& .053 & (0.08, 0.33, 0.51) &&& .061 & (0.08, 0.30, 0.49) &&& .059 & (0.07, 0.33, 0.51)  \\
    & & quad & .061 & (0.05. 0.06, 0.06) &&& .062 & (0.05, 0.06, 0.06) &&& .064 & (0.13, 0.20, 0.58) &&& .069 & (0.09, 0.16, 0.52) \\
    \midrule
    & $(5000, 50)$ & linear & .058 & (0.24, 0.59, 0.86) &&& .054 & (0.23, 0.59, 0.88) &&& .053 & (0.26, 0.62, 0.89) &&& .060 & (0.27, 0.61, 0.88) \\
    & & log & .062 & (0.06, 0.13, 0.11) &&& .051 & (0.17, 0.46, 0.68) &&& .066 & (0.20, 0.64, 0.86) &&& .068 & (0.22, 0.66, 0.84)  \\
    & & cube-root & .053 & (0.16, 0.26, 0.31) &&& .056 & (0.22, 0.61, 0.88) &&& .046 & (0.24, 0.57, 0.88) &&& .042 & (0.26, 0.58, 0.89) \\
    & & inverse & .047 & (0.02, 0.09, 0.03) &&& .040 & (0.03, 0.11, 0.05) &&& .056 & (0.24, 0.58, 0.90) &&& .058 & (0.22, 0.58, 0.87) \\
    & & PL & .058 & (0.21, 0.52, 0.86) &&& .054 & (0.23, 0.55, 0.88) &&& .058 & (0.22, 0.56, 0.89) &&& .054 & (0.22, 0.56, 0.89) \\
    & & quad & .053 & (0.10, 0.09, 0.06) &&& .053 & (0.11, 0.09, 0.06) &&& .043 & (0.24, 0.59, 0.86) &&& .051 & (0.22, 0.52, 0.83) \\
    \midrule
    & $(5000, 100)$ & linear & .052 & (0.15, 0.54, 0.85) &&& .049 & (0.17, 0.56, 0.87) &&& .050 & (0.14, 0.62, 0.89) &&& .056 & (0.17, 0.60, 0.89) \\
    & & log & .044 & (0.07, 0.09, 0.03) &&& .064 & (0.16, 0.32, 0.61) &&& .064 & (0.21, 0.57, 0.86) &&& .069 & (0.25, 0.58, 0.86)  \\
    & & cube-root & .050 & (0.06, 0.14, 0.35) &&& .055 & (0.22, 0.52, 0.85) &&& .047 & (0.21, 0.58, 0.86) &&& .053 & (0.26, 0.56, 0.88) \\
    & & inverse & .053 & (0.07, 0.08, 0.08) &&& .061 & (0.03, 0.07, 0.11) &&& .048 & (0.17, 0.55, 0.88) &&& .063 & (0.18, 0.49, 0.83)  \\
    & & PL & .055 & (0.21, 0.50, 0.80) &&& .058 & (0.25, 0.56, 0.87) &&& .055 & (0.25, 0.59, 0.88) &&& .060 & (0.22, 0.58, 0.85) \\
    & & quad & .056 & (0.07, 0.04, 0.10) &&& .057 & (0.06, 0.04, 0.10) &&& .050 & (0.26, 0.61, 0.87) &&& .051 & (0.23, 0.51, 0.84) \\
    \bottomrule
    \end{tabular}}
    \caption{Empirical Type I error and power of the proposed nonlinear causal test for the simulated example (categorical instrument variables) in Example 3 of Section 3.}
    \label{tab:sim_cate}
\end{table*}

\begin{table*}[!ht]
    \centering
    \scalebox{.75}{
    \begin{tabular}{@{}ccccccccccccccccccc@{}}
    \toprule
    & ~ & ~ & \multicolumn{2}{c}{\textit{2SLS}}  & ~ & ~ & \multicolumn{2}{c}{\textit{PT-2SLS}} & ~ & ~ & \multicolumn{2}{c}{\textit{2SIR} (proposed)} && \\
    & $(n, p)$ & & coverage & length & && coverage & length &&& coverage & length \\
    \midrule
    & $(2000, 10)$ & linear & 0.939 & 0.179 &&& 0.940 & 0.179 &&& 0.980 & 0.198 \\
    & & log & 1.000 & 331.983 &&& 0.936 & 0.184 &&& 0.974 & 0.199 \\
    & & cube-root & 1.000 & 0.694 &&& 0.930 & 0.181 &&& 0.979 & 0.201 \\
    & & inverse & 0.998 & 0.503 &&& 0.810 & 0.130 &&& 0.980 & 0.199 \\
    & & PL & 0.959 & 0.183 &&& 0.957 & 0.184 &&& 0.965 & 0.199 \\
    & & quad & 0.863 & 0.126 &&& 0.834 & 0.122 &&& 0.986 & 0.197 \\
    \midrule
    & $(2000, 50)$ & linear & 0.951 & 0.178 &&& 0.955 & 0.179 &&& 0.975 & 0.197 \\
    & & log & 0.953 & 370.745 &&& 0.937 & 0.158 &&& 0.982 & 0.195 \\
    & & cube-root & 1.000 & 0.610 &&& 0.954 & 0.166 &&& 0.980 & 0.198 \\
    & & inverse & 0.996 & 0.802 &&& 0.922 & 0.134 &&& 0.991 & 0.198 \\
    & & PL & 0.964 & 0.173 &&& 0.953 & 0.175 &&& 0.974 & 0.197 \\
    & & quad & 0.892 & 0.143 &&& 0.893 & 0.140 &&& 0.964 & 0.202 \\
    \midrule
    & $(5000, 50)$ & linear & 0.960 & 0.129 &&& 0.961 & 0.129 &&& 0.979 & 0.133 \\
    & & log & 1.000 & 250.982 &&& 0.910 & 0.112 &&& 0.978 & 0.133 \\
    & & cube-root & 1.000 & 0.312 &&& 0.945 & 0.125 &&& 0.975 & 0.134 \\
    & & inverse & 0.968 & 0.461 &&& 0.781 & 0.085 &&& 0.980 & 0.133 \\
    & & PL & 0.956 & 0.129 &&& 0.961 & 0.130 &&& 0.972 & 0.135 \\
    & & quad & 0.783 & 0.086 &&& 0.785 & 0.085 &&& 0.957 & 0.135 \\
    \midrule
    & $(10000, 50)$ & linear & 0.960 & 0.102 &&& 0.954 & 0.102 &&& 0.977 & 0.104 \\
    & & log & 1.000 & 145.694 &&& 0.964 & 0.101 &&& 0.976 & 0.104 \\
    & & cube-root & 1.000 & 0.218 &&& 0.932 & 0.101 &&& 0.971 & 0.103 \\
    & & inverse & 0.892 & 0.589 &&& 0.670 & 0.060 &&& 0.979 & 0.104 \\
    & & PL & 0.958 & 0.103 &&& 0.956 & 0.103 &&& 0.982 & 0.103 \\
    & & quad & 0.673 & 0.067 &&& 0.666 & 0.066 &&& 0.987 & 0.103 \\
    \bottomrule
    \end{tabular}}
    \caption{Empirical coverage and length of the CI for the simulated example (marginal effect inference) with categorical instrument variables in Example 3 of Section 3. }
    \label{tab:sim_CI_cate}
\end{table*}

\begin{figure}[h!]
    \centering
    \includegraphics[scale=.4]{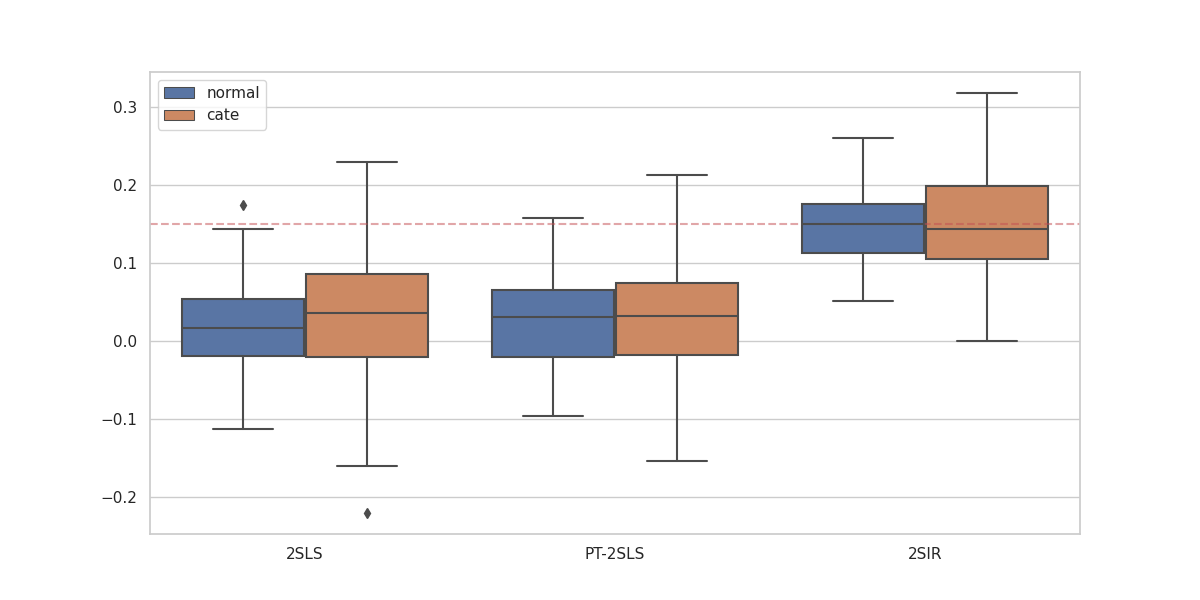}
    \caption{The boxplot for estimated marginal causal effect $\beta$ for both normal distributed and categorical instrument variables based on an ``inverse'' transformation function in Example 3 of Section 3 with $n=2000, p = 10, \beta=.15$.}
    \label{fig:sim_cate}
\end{figure}

\subsection{Simulation results for weak IVs}

\begin{figure}[h!]
    \centering
    \includegraphics[scale=.31]{figs/sim_test_n5p5.png}
    \includegraphics[scale=.31]{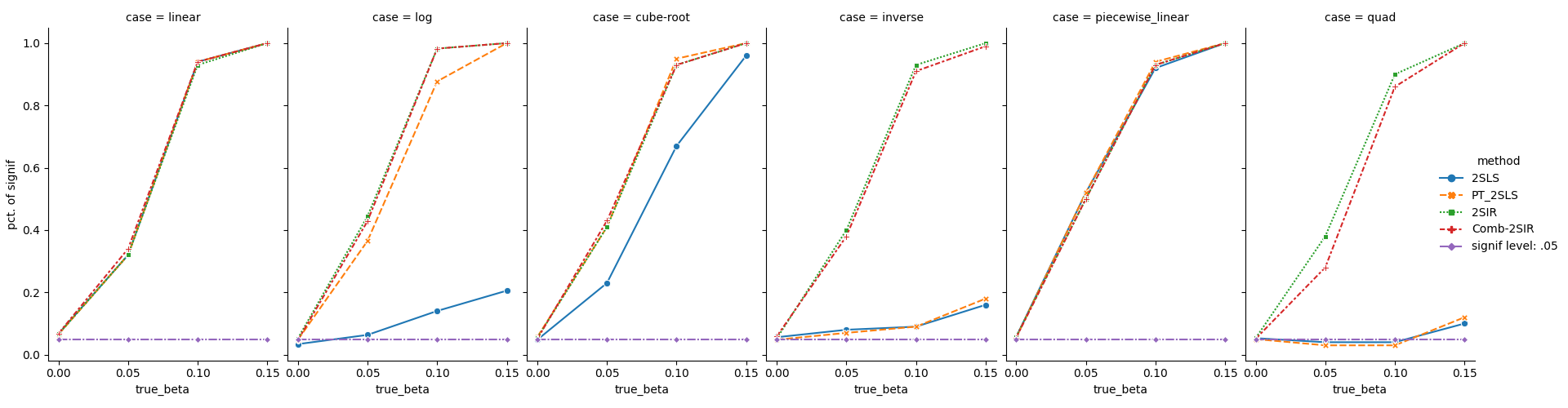}
    \includegraphics[scale=.31]{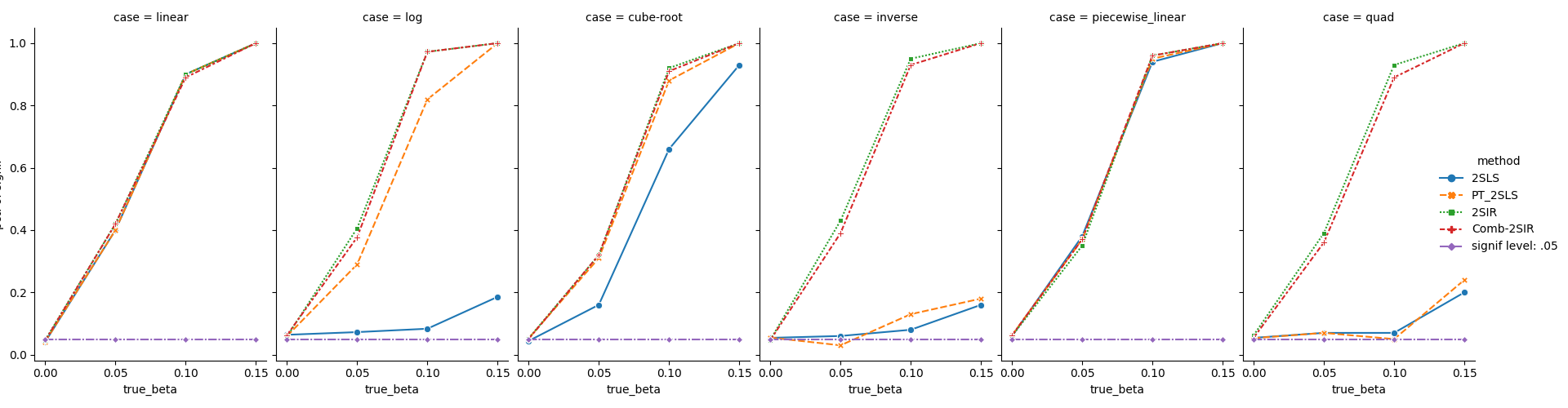}
    \caption{Empirical Type I error (for $\beta_0 = 0$) and power (for $\beta_0 = 0.05, 0.10, 0.15$) of the proposed nonlinear causal test for the simulated example (marginal effect inference) in Example 4 (weak IVs) of Section 3, $\pi = 0.0, 0.1, 0.3$ from up to bottom. Here {2SLS}, {PT-2SLS}, {2SIR}, and {Comb-2SIR} denote two-stage least square, Yeo-Johnson power transformed two-stage least square, the proposed method, and the Cauchy combined proposed method, respectively.}
    \label{fig:sim_weak}
\end{figure}

\noindent {\textbf{Example 4} (Weak IVs). In this example, we examine the performance and stability of the proposed method with weak IVs. 
Specifically, we set $\bm{\theta} \sim N(\bm{0}, \bm{I}_p)$, $\theta_{j} = 0; j = 1, \cdots, \lfloor \pi p \rfloor$, and normalize it by its norm, then $x_i$ and $y_i$ are generated following the same procedure in Example 1 based on $(n=5000, p=50)$, and $\pi = 0.0, 0.1, 0.3$. All empirical results are summarized in Figure \ref{fig:sim_weak} (testing), Table \ref{tab:sim_CI_weak} (CI).}

\begin{table*}[h!]
    \centering
    \scalebox{.75}{
    \begin{tabular}{@{}ccccccccccccccccccc@{}}
    \toprule
    & ~ & ~ & \multicolumn{2}{c}{\textit{2SLS}}  & ~ & ~ & \multicolumn{2}{c}{\textit{PT-2SLS}} & ~ & ~ & \multicolumn{2}{c}{\textit{2SIR} (proposed)} && \\
    & $\pi$ & & coverage & length & && coverage & length &&& coverage & length \\
   \midrule
    & 0.0 & linear & 0.950 & 0.094 &&& 0.952 & 0.095 &&& 0.978 & 0.095 \\
    & & log & 1.000 & 95.559 &&& 1.000 & 0.090 &&& 0.972 & 0.097 \\
    & & cube-root & 1.000 & 0.215 &&& 0.999 & 0.095 &&& 0.982 & 0.097 \\
    & & inverse & 0.801 & 0.209 &&& 0.640 & 0.060 &&& 0.972 & 0.096 \\
    & & PL & 0.951 & 0.096 &&& 0.960 & 0.096 &&& 0.977 & 0.096 \\
    & & quad & 0.522 & 0.052 &&& 0.523 & 0.051 &&& 0.976 & 0.095 \\
    \midrule
    & 0.1 & linear & 0.952 & 0.096 &&& 0.952 & 0.096 &&& 0.972 & 0.095  \\
    & & log & 1.000 & 104.281 &&& 0.947 & 0.090 &&& 0.965 & 0.095  \\
    & & cube-root & 1.000 & 0.206 &&& 0.947 & 0.094 &&& 0.968 & 0.094 \\
    & & inverse & 0.775 & 0.485 &&& 0.607 & 0.059 &&& 0.960 & 0.094 \\
    & & PL & 0.952 & 0.095 &&& 0.955 & 0.095 &&& 0.971 & 0.094 \\
    & & quad & 0.584 & 0.054 &&& 0.578 & 0.054 &&& 0.969 & 0.094 \\
    \midrule
    & 0.3 & linear & 0.945 & 0.096 &&& 0.947 & 0.096 &&& 0.968 & 0.095  \\
    & & log & 1.000 & 118.202 &&& 0.936 & 0.091 &&& 0.955 & 0.095  \\
    & & cube-root & 1.000 & 0.222 &&& 0.958 & 0.097 &&& 0.966 & 0.096 \\
    & & inverse & 0.775 & 1.026 &&& 0.597 & 0.054 &&& 0.964 & 0.094 \\
    & & PL & 0.936 & 0.095 &&& 0.943 & 0.095 &&& 0.971 & 0.094 \\
    & & quad & 0.566 & 0.054 &&& 0.567 & 0.054 &&& 0.975 & 0.094 \\
    \bottomrule
    \end{tabular}}
    \caption{Empirical coverage and length of the confidence interval for the simulated example (marginal effect inference)  in Example 4 (weak IVs) of Section 3.}
    \label{tab:sim_CI_weak}
\end{table*}

\subsection{Simulation results for non-additive and epistatic effects}

\begin{figure}[h]
    \centering
    \includegraphics[scale=.3]{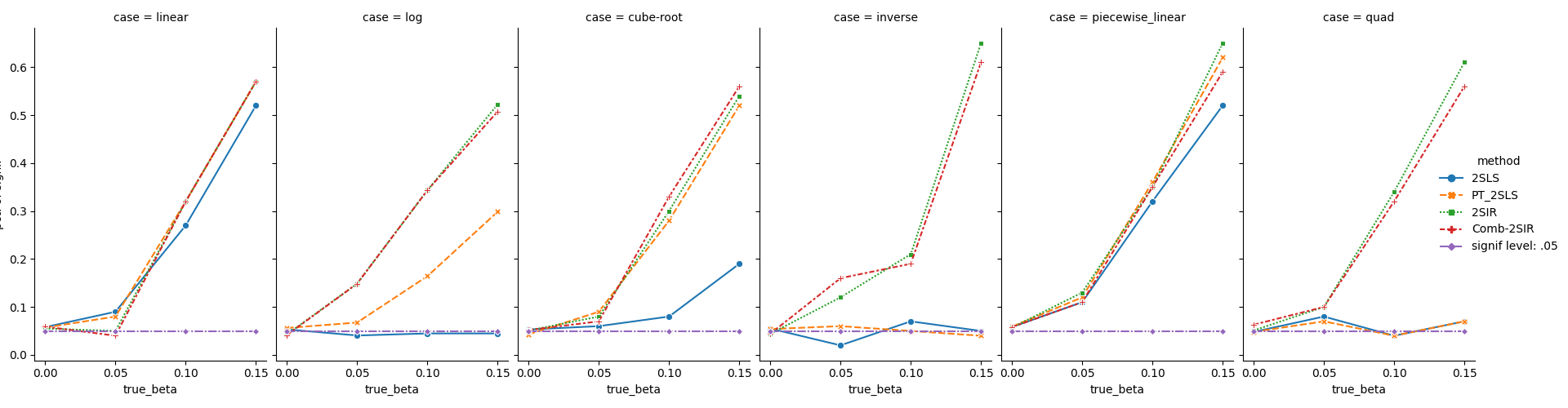}
    \includegraphics[scale=.3]{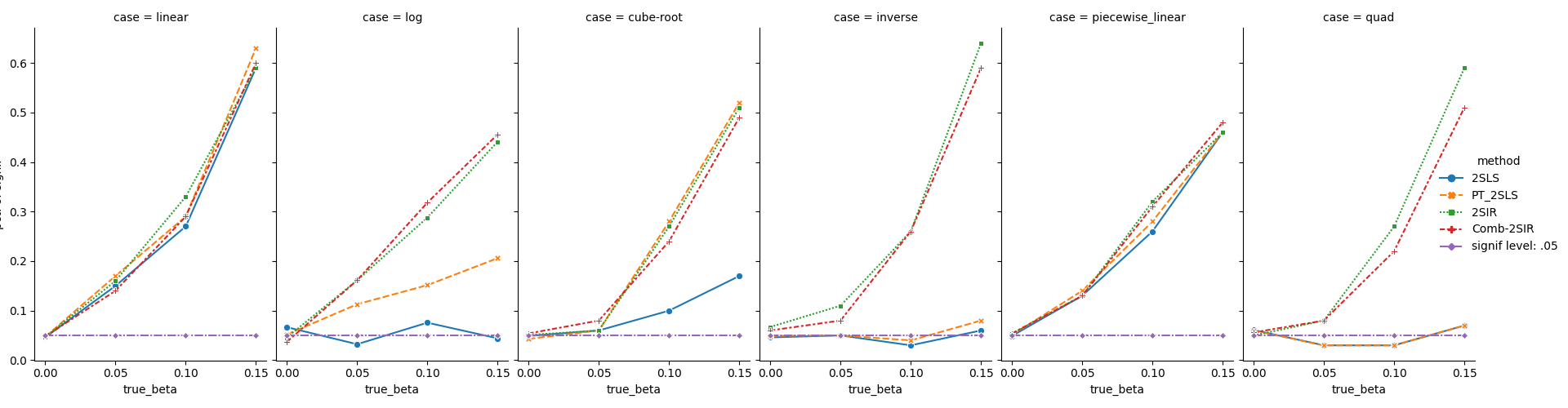}
    \includegraphics[scale=.3]{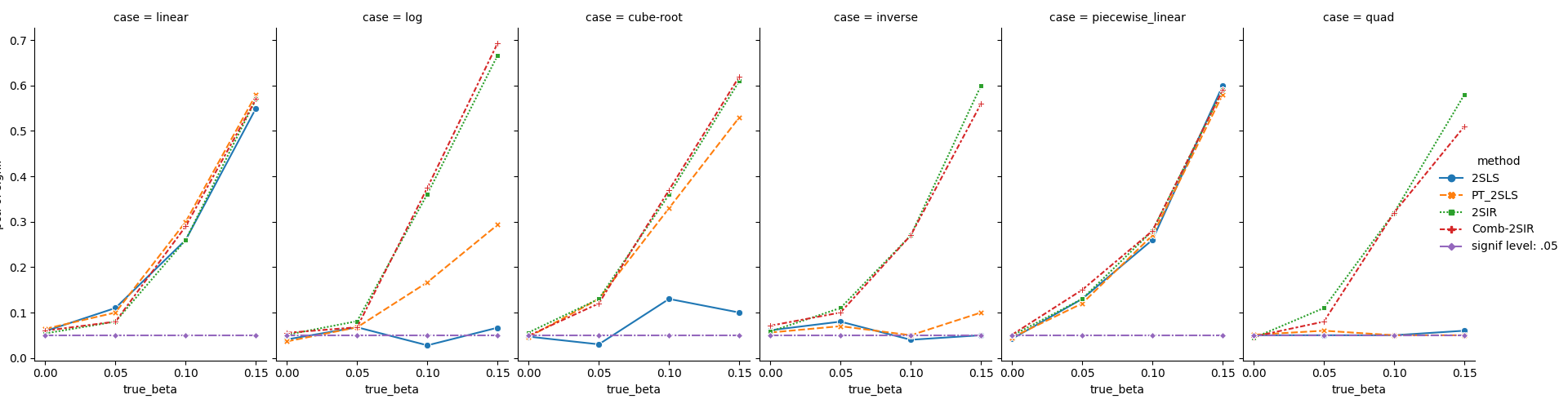}
    \includegraphics[scale=.3]{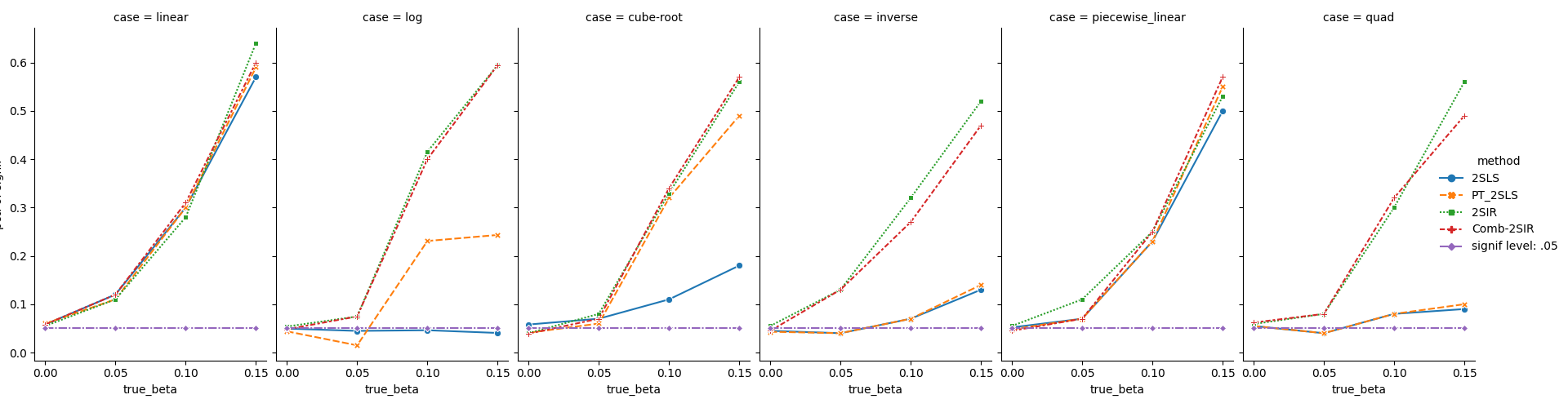}
    \caption{Empirical Type I error (for $\beta_0 = 0$) and power (for $\beta_0 = 0.05, 0.10, 0.15$) of the proposed nonlinear causal test for the simulated example (marginal effect inference) in Example 5 (non-additive and epistatic effects) of Section 3, $(\lambda, |\mathcal{J}|) = (0.3, \lfloor 0.1p \rfloor), (0.5, \lfloor 0.1p \rfloor), (0.3, \lfloor 0.3p \rfloor), (0.5, \lfloor 0.3p \rfloor)$ from up to bottom.}
    \label{fig:sim_DE}
\end{figure}

\noindent {\textbf{Example 5} (Non-additive and epistatic effects). In this example, we examine the performance and stability of the proposed method under non-additive and epistatic genetic effects. First, $(\bm{z}_i)_{i=1, \cdots, n}$ are generated based on the same setting in Example 3. To incorporate the non-additive and epistatic effects, $x_i = \phi^{-1}( \bm{\theta}^{\intercal}_a I(\bm{z}_i = 1) + \bm{\theta}^{\intercal}_d I(\bm{z}_i = 2) + \sum_{(j,j') \in \mathcal{J}} \delta_{j,j'} z_{ij} z_{ij'} + w_i)$. Here, we set $\bm{\theta}_a \sim N(\bm{0}, \bm{I}_p)$, and $\bm{\theta}_d = \lambda \bm{\theta}_a$ presents non-additive effects when $\lambda \neq 2$. Besides, $\bm{\delta} \sim N(\bm{0}, 0.1\bm{I}_{|\mathcal{J}|})$ presents epistatic (i.e. interaction) effects, and $\mathcal{J}$ is a set of randomly selected pairs, where each pair is uniformly sampled. Finally, $y_i$ is generated following the same procedure in Example 3. In this example, we set $n = 5000, p = 50$, $\lambda = 0.3, 0.5$ and $|\mathcal{J}| = \lfloor 0.1p \rfloor, \lfloor 0.3p \rfloor$. All empirical results are summarized in Figure \ref{fig:sim_DE} (testing), Table \ref{tab:sim_CI_DE} (CI).}

\begin{table*}[h!]
    \centering
    \scalebox{.7}{
    \begin{tabular}{@{}ccccccccccccccccccc@{}}
    \toprule
    & ~ & ~ & \multicolumn{2}{c}{\textit{2SLS}}  & ~ & ~ & \multicolumn{2}{c}{\textit{PT-2SLS}} & ~ & ~ & \multicolumn{2}{c}{\textit{2SIR} (proposed)} && \\
    & $(\lambda, |\mathcal{J}|)$ & & coverage & length & && coverage & length &&& coverage & length \\
    \midrule
    & (1.3, 0.1$p$) & linear & 0.945 & 0.115 &&& 0.942 & 0.116 &&& 0.992 & 0.123 \\
    & & log & 1.000 & 221.321 &&& 0.889 & 0.103 &&& 0.989 & 0.124 \\
    & & cube-root & 1.000 & 0.281 &&& 0.915 & 0.115 &&& 0.987 & 0.125 \\
    & & inverse & 0.979 & 0.331 &&& 0.796 & 0.087 &&& 0.992 & 0.123 \\
    & & PL & 0.937 & 0.115 &&& 0.932 & 0.116 &&& 0.986 & 0.125 \\
    & & quad & 0.787 & 0.084 &&& 0.780 & 0.084 &&& 0.998 & 0.125 \\
    \midrule
    & (1.3, 0.3$p$) & linear & 0.923 & 0.113 &&& 0.925 & 0.113 &&& 0.989 & 0.123 \\
    & & log & 1.000 & 234.423 &&& 0.892 & 0.103 &&& 0.992 & 0.124 \\
    & & cube-root & 1.000 & 0.275 &&& 0.907 & 0.112 &&& 0.989 & 0.123 \\
    & & inverse & 0.977 & 1.920 &&& 0.772 & 0.087 &&& 0.982 & 0.123 \\
    & & PL & 0.918 & 0.112 &&& 0.914 & 0.113 &&& 0.985 & 0.123 \\
    & & quad & 0.785 & 0.084 &&& 0.784 & 0.083 &&& 0.990 & 0.123 \\
   \midrule
    & (1.5, 0.1$p$) & linear & 0.948 & 0.115 &&& 0.952 & 0.116 &&& 0.992 & 0.124 \\
    & & log & 1.000 & 188.532 &&& 0.893 & 0.105 &&& 0.991 & 0.121 \\
    & & cube-root & 1.000 & 0.268 &&& 0.917 & 0.112 &&& 0.982 & 0.123 \\
    & & inverse & 0.977 & 0.334 &&& 0.801 & 0.088 &&& 0.989 & 0.124 \\
    & & PL & 0.944 & 0.115 &&& 0.945 & 0.116 &&& 0.991 & 0.124 \\
    & & quad & 0.776 & 0.082 &&& 0.773 & 0.082 &&& 0.988 & 0.124 \\
    \midrule
    & (1.5, 0.3$p$) & linear & 0.931 & 0.115 &&& 0.942 & 0.114 &&& 0.989 & 0.123 \\
    & & log & 1.000 & 210.321 &&& 0.882 & 0.104 &&& 0.991 & 0.123 \\
    & & cube-root & 1.000 & 0.281 &&& 0.942 & 0.114 &&& 0.995 & 0.124 \\
    & & inverse & 0.974 & 0.583 &&& 0.799 & 0.085 &&& 0.986 & 0.124 \\
    & & PL & 0.926 & 0.114 &&& 0.928 & 0.115 &&& 0.986 & 0.123 \\
    & & quad & 0.789 & 0.085 &&& 0.788 & 0.085 &&& 0.990 & 0.123 \\
    \bottomrule
    \end{tabular}}
    \caption{Empirical coverage and length of the CI for the simulated example (marginal effect inference) in Example 5 (non-additive and epistatic effects) of Section 3.}
    \label{tab:sim_CI_DE}
\end{table*}

\subsection{Simulation results for misspecified models}
\label{sec:sim_misspecified}

\begin{figure}[h]
    \centering
    \includegraphics[scale=.31]{figs/sim_test_mm.png}
    \includegraphics[scale=.31]{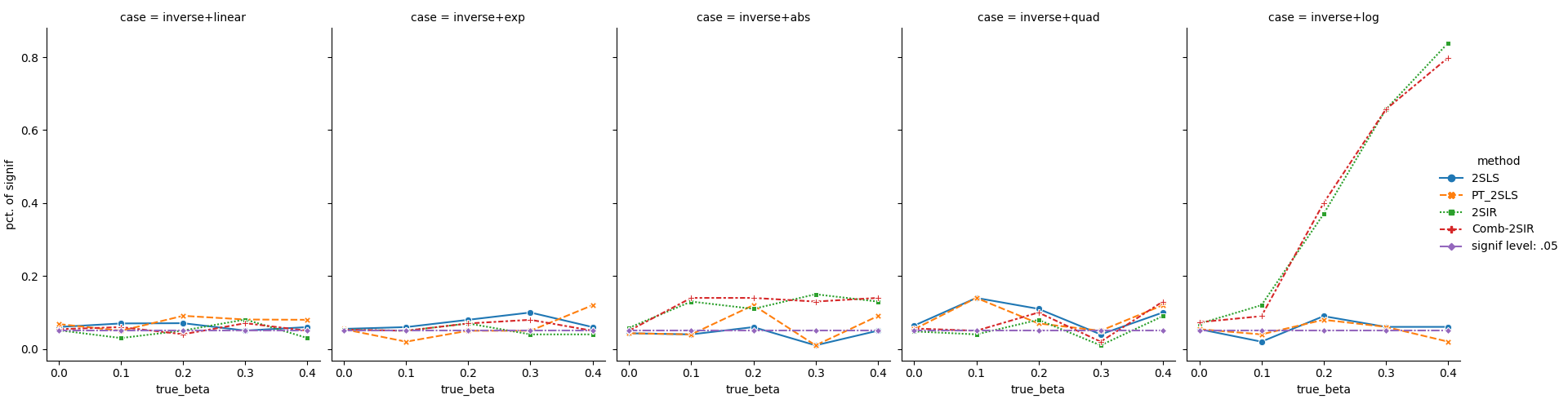}
    \caption{Empirical Type I error ($\beta_0 = 0$) and power ($\beta_0 = 0.1, 0.2, 0.3, 0.4$) of the proposed nonlinear causal test for the simulated example with misspecified causal transformation in Example 6 of Section 3. Here $\phi(x) = x^2$ and $\phi(x) = 1/x$ are specified for two rows, respectively; and $\psi(x) = x$, $\psi(x) = e^x$, $\psi(x) = |x|$, $\psi(x) = 1/x$, $\psi(x)=\log(|x|)$ are specified for five columns.}
    \label{fig:sim_mm}
\end{figure}

\noindent {\textbf{Example 6} (Misspecified models). We examine the performance and stability of the proposed method for misspecified models. Specifically, $(\bm{z}_i, x_i)_{i=1, \cdots, n}$ are generated with the same procedure in Example 1. In Stage 2, we consider misspecified models: $y_i = \beta \psi(x_i) + \epsilon_i$ with $\psi(x) = x$, $\psi(x) = \exp(x)$, $\psi(x) = |x|$, $\psi(x) = 1/x$, and $\psi(x) = \log(|x|)$. According to the simulation results in Example 1, we mainly consider $\phi(x) = x^2$ and $\phi(x) = 1/x$ to highlight the differences between the proposed methods and other competitors. All empirical results are summarized in Figure \ref{fig:sim_mm} (testing).} 

\subsection{R-squared values for the estimated equation}
This subsection includes the R-squared values for the estimated equation ($\mathbf{z}$-$x$) based on the ADNI dataset. The numerical results are summarized in the folder "app\_S11\_r2".

\section{Supplementary results and technical proofs}

\subsection{Selection bias}

{
In the real data application, we pre-screen SNPs based on multiple criteria. 
This subsection analyzes the potential selection bias in our procedure. 
To this end, consider the following situation. 
Suppose $(\bm z, x, y)$ comes from the model 
\begin{equation*}
    \begin{split}
        \phi(x) = \bm z^\intercal \bm\theta + w, \qquad 
        y = \beta\phi(x) + \bm z^\intercal \bm\alpha + \varepsilon,
    \end{split}
\end{equation*}
where $\bm z\in\mathbb{R}^d$, $\bm\theta = (\bm\theta_M,\bm 0)$, $|M|\ll d$, and the other settings remain the same as model (1).
The prescreening procedure  based on $(\bm Z_1, \bm X_1)\in\mathbb{R}^{n_1\times(d + 1)}$ selects a model $\widehat{M}$ with cardinality $|\widehat{M}|= p\ll d$ being fixed.
Assume the prescreening procedure satisfies the sure screening property \citep{fan2008sure} in that 
$P(\widehat{M}\supseteq M) \to 1$. Moreover, $A = \{ j: \alpha_j\neq 0 \}\subseteq M$. 
For any $M'\supseteq M$ with $|M'|=p$, we have a submodel
\begin{equation*}
    \begin{split}
        \phi(x) = \bm z^\intercal_{M'} \bm\theta^{(M')} + w, \qquad
        y = \beta\phi(x) + \bm z^\intercal_{M'} \bm\alpha^{(M')} + \varepsilon,
    \end{split}
\end{equation*}
where $\bm\theta^{(M')} = (\bm \theta_M,\bm 0)\in\mathbb{R}^{|M'|}$ and $\bm\alpha^{(M')} = (\bm \alpha_{A},\bm 0)\in\mathbb{R}^{|M'|}$.
Let $\widehat{\bm\theta}^{(M')}$ be the SIR estimator based on $M'$. Then by Theorem 4 of \citep{zhu1995asymptotics}, 
we have $\sqrt{n_1}(\widehat{\bm\theta}^{(M')} - \bm\theta^{(M')})\dto \bm\xi^{(M')}$ for a subgaussian random variable $\bm\xi^{(M')}$. 
Since there are $\binom{d}{p-|M|} \leq (ed/ (p-|M|))^{p-|M|}$ possible $M'$, we have 
\begin{equation*}
    \max_{M'\supseteq M:|M'|=p} \sqrt{n_1}|\widehat{\bm\theta}^{(M')} - \bm\theta^{(M')}| = O_p\Big(\sqrt{ (p-|M|)\log(d) }\Big).
\end{equation*}
Thus, $\widehat{\bm\theta} = \widehat{\bm\theta}^{(\widehat{M})}$ is consistent provided that $n_1\gg (p-|M|)\log(d)$.
It follows that $\widehat{\beta}$ is also consistent in this situation. 
In view of Theorem 1, $\sqrt{n_2}\widehat{\beta}\dto |N(0,(\bm\theta^\intercal\widetilde{\bm\Sigma}\bm\theta)^{-1}\sigma_e^2)|$ when $\beta = 0$.
Consequently, the test (8) of $H_0:\beta = 0$ remains valid after a sure screening procedure. 
To conclude, our procedure seems largely immune to the potential selection bias provided that the sample size $n_1\gg (p-|M|)\log(d)$ and a sure screening method is used.
}

\subsection{Regularity conditions and supplementary results}
\label{sec:regularity-condition}

We impose the following regularity conditions for 2SIR and AIR.
In particular, Condition \ref{condition:sir} is used to establish the asymptotic distribution of 
SIR estimate $\widehat{\bm\theta}$, Conditions \ref{condition:sir} and \ref{condition:sparse-regression} are used to derive the asymptotic properties of 
2SIR estimate $\widehat{\beta}$, and Conditions \ref{condition:sir} and \ref{condition:air} are 
used to quantify the convergence rate of AIR estimate $\widehat{\phi}$.

\begin{condition}\label{condition:sir}
    Assume the following conditions for sliced inverse regression.\\
    \indent (i) $\E(\bm z\mid \bm z^\intercal\bm\theta)$ is linear in $\bm z^\intercal\bm\theta$;\\
    \indent (ii) $c_{-}\leq \lambda_{\min}(\bm \Sigma)\leq \lambda_{\max}(\bm \Sigma)\leq c_+$, where $\bm \Sigma = \E \bm z \bm z^\intercal$;\\
    \indent (iii) $\E\|\bm z\|^4 <\infty$; \\ 
    \indent (iv) 
    $\E(\bm z\mid x)$ has a total variation of order $1/4$ in that 
    \begin{equation*}
        \lim_{n_1\to\infty} 
        \frac{1}{n_1^{1/4}} \sup_{\Pi_{n_1}(D)} \sum_{i=1}^{n_1-1}
        \|\E(\bm z\mid x^*_{(i+1)})-\E(\bm z\mid x^*_{(i)})\| = 0,
    \end{equation*} 
    where $\Pi_{n_1}(D)$ is the collection of all $n_1$-point partitions,
    $-D\leq x^*_{(1)}\leq \cdots\leq x^*_{(n_1)}\leq D$ of the interval $[-D,D]$,
    $D>0$ and $\|\cdot\|$ is the Euclidean norm;\\
    \indent (v) There exist a nondecreasing real-valued function $M$ and a real number $D_0>0$ such that for any two points $x_1,x_2 < -D_0$
    or $x_1,x_2>D_0$,
    \begin{equation*}
        \|\E(\bm z\mid x_1)-\E(\bm z\mid x_2)\|\leq |M(x_1)-M(x_2)|,
    \end{equation*}
    and $M^4(t) P(x>t) \to 0$ as $t\to\infty$, as $n_1\to\infty$;\\
    \indent (vi) 
    Let $\Cov(\bm u\mid x)$ has a total variation of order $1$ in that 
    \begin{equation*}
        \lim_{n_1\to\infty} 
        \frac{1}{n_1} \sup_{\Pi_{n_1}(D)} \sum_{i=1}^{n_1-1}
        \|\Cov(\bm u\mid x^*_{(i+1)})-\Cov(\bm u\mid x^*_{(i)})\|_F = 0,
    \end{equation*}
    where $\bm u = \bm z - \E(\bm z \mid x)$ and $\|\cdot\|_F$ is the Frobenius norm.
\end{condition}

Condition \ref{condition:sir} is common in the sufficient dimension reduction literature \citep{zhu1995asymptotics,zhu2006sliced}. 
Note that (i) and (ii) impose distributional assumptions on $\bm z$, where (i) is equivalent to that $\bm z$ has an elliptically symmetric distribution \citep{cook1991}. However, it can be approximately extended to categorical IVs as indicated in \citep{peter1993}. Moreover, the numerical performance in Example 3 of Section 3 also suggests that the proposed method can apply to categorical IVs. 
Condition \ref{condition:sir} (iii)-(vi) are used to derive the asymptotic distribution of $\widehat{\bm\theta}$; see \citep{zhu1995asymptotics} for details. 
Under Condition 1, we have 
$n_1^{1/2}( \widehat{\bm\theta}-\bm\theta) \dto \bm\xi$,
where the distribution of $\bm\xi$ is given in Theorem 4 of \citep{zhu1995asymptotics}. 

{For estimating $\beta$, we aim to solve a sparse regression problem in (4) of the main text. 
In (4), $\|\cdot\|_0$ penalty is used. For theoretical analysis, we also consider its surrogates SCAD, TLP, and MCP, defined as follows:
\begin{itemize}
    \item (SCAD)
    \begin{equation}\label{eqn:SCAD}
        \begin{split}
            p_{a}(t) = 
        \begin{cases}
            4|t|/3a & |t| \leq a/2,\\
            1 - 4(|t| - a)^2/3a^2  & a/2 < |t| \leq a \\
            1 & |t| > a,
        \end{cases}
        \end{split}
    \end{equation}
    \item (TLP)
    \begin{equation}\label{eqn:TLP}
        \begin{split}
            p_{a}(t) = 
        \begin{cases}
            |t|/a & |t| \leq a,\\
            1 & |t| > a,
        \end{cases} 
        \end{split}
    \end{equation}
    \item (MCP)
    \begin{equation}\label{eqn:MCP}
        \begin{split}
            p_{a}(t) = 
        \begin{cases}
            2|t|/a - |t|^2/a^2 & |t| \leq a,\\
            1 & |t| > a,
        \end{cases}
        \end{split}
    \end{equation}
\end{itemize}
where $a > 0$ is a hyperparameter.}

\begin{condition}\label{condition:sparse-regression}
    Assume the following conditions are satisfied.\\
    \indent (i) $|A| < p/2$, where $A=\{ j : \alpha_j \neq 0 \}$;\\
    \indent (ii) $\|\bm\alpha\|_{\min} = \min_{j\in A}|\alpha_j|\geq 32 \sigma_e \sqrt{c_+} c_{-}^{-1} \sqrt{\log(n)/n}$;\\
    \indent (iii) $n_2/n_1\to r\in (0,\infty)$. \\
    \indent {(iv) $0 < a < \sqrt{c_+\log(n)/(2p\lambda_{\max}(\bm Z_2^\intercal\bm Z_2))}$ when SCAD, TLP, or MCP is used.}
\end{condition}



Condition \ref{condition:sparse-regression} (i) is an assumption for identifiability of $\beta$, cf. Corollary 1 of \citep{kang2016instrumental},
while (ii) is nearly necessary for the consistent selection of invalid instruments \citep{wainwright2009information}. 
Condition \ref{condition:sparse-regression} (iii) is a common assumption in two-sample inference \citep{pacini2016robust}. 

For the estimation of nonlinear transformation $\phi$, we impose the following condition.

\begin{condition}\label{condition:air}
    Assume $\widehat m$ satisfies the following properties.\\
    \indent (i) $\E\|\widehat{m}-m\|_{\infty}\leq c_1 n_1^{-\kappa_1}$, where $\|m\|_\infty = \sup_{x\in\mathcal X}|m(x)|$ and $\kappa_1>0$;\\
    \indent (ii) $\E n_1^{-1} \sum_{i=1}^{n_1}|\widehat{m}(x_{1i})-m(x_{1i})|^2\leq c_2 {n}_1^{-\kappa_2}$, where $\kappa_2>0$.
\end{condition}

In Condition \ref{condition:air}, 
(i) specifies the local estimation quality via the sup-norm convergence rate over the treatment region of interest $\mathcal X$,
while (ii) specifies the global estimation quality 
via the convergence rate in the empirical $L_2$-norm.  
The convergence results of various nonparametric regressions have been extensively studied; see \citep{tsybakov2008introduction} for an overview. 

Theorem \ref{theorem:air} presents the convergence rate for estimating nonlinear transformation $\phi(\cdot)$ and nonlinear causal effect $\beta\phi(\cdot)$.

\begin{theorem}\label{theorem:air}
    Assume Conditions \ref{condition:sir} and \ref{condition:air} in Section \ref{sec:regularity-condition} and conditions in Proposition 1. Then 
    $\|\widehat{\phi}-\phi\|_{\infty} \leq O_p(\max(\sqrt{p/n_1}, n_1^{-\min(\kappa_1,\kappa_2)}))$.
    If in addition Condition \ref{condition:sparse-regression} in Section \ref{sec:regularity-condition} holds,
    then $\|\widehat{\beta}\widehat{\phi} - \beta\phi\|_{\infty} = O_p(\max( \sqrt{p/n_1}, \sqrt{1/n_2}, n_1^{-\min(\kappa_1,\kappa_2)}))$, 
    where  $\|\cdot\|_{\infty}$ is sup-norm given in Condition \ref{condition:air}, and $\kappa_1,\kappa_2>0$ are convergence rates of $\widehat{m}$ in (5).
\end{theorem}

Theorem \ref{theorem:air} shows that the convergence rate of 
$\widehat\beta\widehat\phi$ is determined by the slowest rate of estimating $\bm\theta$, $\beta$, and $m(\cdot)$. Note that the estimation of $\bm\theta$ and $\beta$ possesses a parametric root-$n$ rate. Hence, the overall convergence rate $\|\widehat{\beta}\widehat{\phi} - \beta\phi\|_{\infty}$ is usually determined by that of the nonparametric function estimation.

\subsection{Technical proofs}

\begin{proof}[Proof of Proposition 1]
    Note that $\E(\bm z^\intercal\bm\theta\mid x) = \E(\bm z^\intercal\bm\theta\mid \phi(x)) = \E(\bm z^\intercal\bm\theta\mid \bm z^\intercal\bm\theta + w)$.
    By the property of elliptical symmetry, $\E(\bm z^\intercal\bm\theta\mid \bm z^\intercal\bm\theta+w)=\rho(\bm z^\intercal\bm\theta + w) = \rho\phi(x)$.
\end{proof}

\begin{lemma}[Theorem 1 of \citep{zhu1995asymptotics}]\label{lemma:sir}
    Assume Condition \ref{condition:sir} is satisfied, then we have 
    $n_1^{-1/2}(\widehat{\bm\theta}-\bm\theta) \dto \bm\xi$, 
    where the distribution of $\bm\xi$ is given in Theorem 1 of \citep{zhu1995asymptotics}. 
\end{lemma}

\begin{lemma}\label{lemma:sparse-regression}
    Under Condition \ref{condition:sparse-regression}, 
    if $K = |A|$, then $P(\widehat{A} \neq A) \leq 4 n_2^{-3}$,
    where $\widehat{A}=\{ j : \widehat{\alpha}_j\neq 0 \}$.
\end{lemma}
\begin{proof}[Proof of Lemma \ref{lemma:sparse-regression}]
    Denote 
    $\widehat{\bm X}_2 = (\bm{z}_{21}^\intercal\widehat{\bm\theta},\ldots,\bm{z}_{2n_2}^\intercal\widehat{\bm\theta})^\intercal$,
    and let $\widetilde{\bm Z} = (\bm Z_2,\widehat{\bm X}_2)\in \mathbb{R}^{n_2\times(p+1)}$ be the augmented data matrix 
    and $\widehat{\bm\gamma}^\circ = (\widehat{\bm\alpha}^\circ,\widehat{\beta}^\circ)$ be the oracle estimator. 
    Let $B = \{j: \widehat{\gamma}^\circ_j\neq 0 \} = A \cup \{ p+1 \}$ and $\widehat{B} = \{ j : \widehat{\gamma}_j\neq 0 \} = \widehat{A}\cup \{ p +1 \}$.
    
    {First, suppose $\|\cdot\|_0$ penalty is used in (4).}
    Since $\widehat{\bm\gamma}$ 
    is the solution of (4),
    we have ${n}_2^{-1}\|\bm Y - \widetilde{\bm Z}\widehat{\bm\gamma} \|_2^2
    \leq n_2^{-1}\|\bm Y - \widetilde{\bm Z}\widehat{\bm\gamma}^\circ\|_2^2$,
    which, after rearrangement, yields that 
    \begin{equation*}
        \frac{1}{n_2}\|\widetilde{\bm Z}(\widehat{\bm\gamma}-\widehat{\bm\gamma}^\circ)\|_2^2
        \leq \frac{2}{n_2} \widehat{\bm e}^\intercal \widetilde{\bm Z}(\widehat{\bm\gamma} - \widehat{\bm\gamma}^\circ),
    \end{equation*}
    where $\widehat{\bm e} = \bm Y - \widetilde{\bm Z}\widehat{\bm\gamma}^\circ$ 
    is the residual vector of the oracle estimator. 
    By the first-order optimality condition of the oracle estimator $\widehat{\bm\gamma}^\circ$, we have 
    $\widehat{\bm e}^\intercal \widetilde{\bm Z}_{B} = \bm 0$. Moreover, we have 
    $\widehat{\bm\gamma}_{(\widehat B\cup B)^c} - \widehat{\bm\gamma}^\circ_{(\widehat B\cup B )^c}=\bm 0$. 
    Hence, we have  
    \begin{equation}\label{eqn:l0-bound}
        \begin{split}
            \frac{1}{n_2}\|\widetilde{\bm Z}_{\widehat{B}\cup B}(\widehat{\bm\gamma}_{\widehat{B}\cup B}-\widehat{\bm\gamma}^\circ_{\widehat{B}\cup B})\|_2^2
        &\leq \frac{2}{n_2} \widehat{\bm e}^\intercal \widetilde{\bm Z}_{\widehat{B}\setminus B} (\widehat{\bm\gamma}_{\widehat{B}\setminus B} - \widehat{\bm\gamma}^\circ_{\widehat{B}\setminus B})\\
        &\leq \frac{2}{n_2} \|\widehat{\bm e}^\intercal \widetilde{\bm Z}_{\widehat{B}\setminus B}\|_2\| \widehat{\bm\gamma}_{\widehat{B}\setminus B} - \widehat{\bm\gamma}^\circ_{\widehat{B}\setminus B}\|_2.
        \end{split}
    \end{equation}
    Further,
    \begin{equation*}
        \begin{split}
            \frac{1}{n_2}\|\widetilde{\bm Z}_{\widehat{B}\cup B}(\widehat{\bm\gamma}_{\widehat{B}\cup B} - \widehat{\bm\gamma}^\circ_{\widehat{B}\cup B})\|_2^2 
        &\geq \lambda_{\min}(n_2^{-1}\widetilde{\bm Z}^\intercal_{\widehat{B}\cup B}\widetilde{\bm Z}_{\widehat{B}\cup B})\|\widehat{\bm\gamma}_{\widehat{B}\cup B} - \widehat{\bm\gamma}^\circ_{\widehat{B}\cup B}\|_2^2.
        \end{split}
    \end{equation*}
    Note that $\|\widehat{\bm\gamma}_{\widehat B\cup B} - \widehat{\bm\gamma}^\circ_{\widehat{B}\cup B}\|_2 \geq  \sqrt{|\widehat{B}\setminus B|}\|\bm\alpha\|_{\min}$
    and $\|\widehat{\bm\gamma}_{\widehat B\cup B} - \widehat{\bm\gamma}^\circ_{\widehat{B}\cup B}\|_2 \geq \| \widehat{\bm\gamma}_{\widehat{B}\setminus B} - \widehat{\bm\gamma}^\circ_{\widehat{B}\setminus B}\|_2$. 
    Combining the above results, we obtain 
    \begin{equation*}
        \begin{split}
        \lambda_{\min}(n_2^{-1}\widetilde{\bm Z}^\intercal_{\widehat{B}\cup B}\widetilde{\bm Z}_{\widehat{B}\cup B}) \sqrt{|\widehat{B}\setminus B|}\|\bm\alpha\|_{\min}
        &\leq \frac{2}{n} \sup_{\{S:|S\setminus B|=|\widehat{B}\setminus B|\}} \|\widehat{\bm e}^\intercal \widetilde{\bm Z}_{S\setminus B}\|_2.
        \end{split}
    \end{equation*}

    Now, let $\mathcal{E} = \mathcal{E}_1\cap \mathcal E_2$, where 
    \begin{equation*}
        \begin{split}
        \mathcal{E}_1 &= \left\{ \lambda_{\min}(n^{-1} \widetilde{\bm Z}^\intercal_{\widehat{B}\cup B} \widetilde{\bm Z}_{\widehat{B}\cup B}) > \frac{c_{-}}{2} \right\},\\
        \mathcal{E}_2 &= \left\{  \sup_{\{ S : |S\setminus B|=k \}} \|\widehat{\bm e}^\intercal \widetilde{\bm Z}_{S\setminus B}\|_2
        \leq 4\sqrt{c_+}\sigma_e \sqrt{k\log(n) / n}, \ 1\leq k\leq |B| \right\}.
        \end{split}
    \end{equation*}
    Then on event $\mathcal{E}$, we have 
    $2^{-1}c_{-}\sqrt{|\widehat{B}\setminus B|}\|\bm\alpha\|_{\min} < 8\sigma_e\sqrt{c_+}\sqrt{|\widehat{B}\setminus B|} \sqrt{\log(n)/n}$. 
    However, by Condition \ref{condition:sparse-regression} (iii), we have 
    $\|\bm\alpha\|_{\min}\geq 32 \sigma_e \sqrt{c_+} c_{-}^{-1} \sqrt{\log(n)/n}$. 
    This implies that $|\widehat{A}\setminus A|=|\widehat{B}\setminus B|=0$, and hence $\widehat A = A$ on event $\mathcal{E}$.
    
    {Next, suppose a surrogate penalty (SCAD, TLP, or MCP) is used in (4). Let $\widehat{A}^*_1 = \{ j : |\widehat{\alpha}_j|>a \}$
    and $\widehat{A}^*_2 = \{ j : |\widehat{\alpha}_j|\leq a \}$. 
    Then \eqref{eqn:l0-bound} needs a modification, 
    \begin{equation}
        \begin{split}
            \frac{1}{n_2}\|\widetilde{\bm Z}_{\widehat{B}\cup B}(\widehat{\bm\gamma}_{\widehat{B}\cup B}-\widehat{\bm\gamma}^\circ_{\widehat{B}\cup B})\|_2^2
        &\leq \frac{2}{n_2} \widehat{\bm e}^\intercal \widetilde{\bm Z}_{\widehat{A}^*_1\setminus A} (\widehat{\bm\gamma}_{\widehat{A}^*_1\setminus A} - \widehat{\bm\gamma}^\circ_{\widehat{A}^*_1\setminus A}) \\ 
        &\quad + \frac{2}{n_2} \widehat{\bm e}^\intercal \widetilde{\bm Z}_{\widehat{A}^*_2\setminus A} (\widehat{\bm\gamma}_{\widehat{A}^*_2\setminus A} - \widehat{\bm\gamma}^\circ_{\widehat{A}^*_2\setminus A}) \\
        &\leq \frac{2}{n_2} \|\widehat{\bm e}^\intercal \widetilde{\bm Z}_{\widehat{A}^*_1\setminus A}\|_2\| \widehat{\bm\gamma}_{\widehat{A}^*_1\setminus A} - \widehat{\bm\gamma}^\circ_{\widehat{A}^*_1\setminus A}\|_2 \\
        &\quad + \frac{2}{n_2} \|\widehat{\bm e}\|_2 \|\widetilde{\bm Z}_{\widehat{A}^*_2\setminus A} (\widehat{\bm\gamma}_{\widehat{A}^*_2\setminus A} - \widehat{\bm\gamma}^\circ_{\widehat{A}^*_2\setminus A})\|_2\\
        &\leq \frac{2}{n_2} \|\widehat{\bm e}^\intercal \widetilde{\bm Z}_{\widehat{A}^*_1\setminus A}\|_2\| \widehat{\bm\gamma}_{\widehat{A}^*_1\setminus A} - \widehat{\bm\gamma}^\circ_{\widehat{A}^*_1\setminus A}\|_2 \\
        &\quad + \frac{3 \sigma_e}{\sqrt{n_2}}  \sqrt{p\lambda_{\max}(\bm Z^\intercal\bm Z)} a.
        \end{split}
    \end{equation}
    We have $\|\widehat{\bm\gamma}_{\widehat B\cup B} - \widehat{\bm\gamma}^\circ_{\widehat{B}\cup B}\|_2 \geq  \sqrt{|\widehat{A}^*_1\setminus A|}\|\bm\alpha\|_{\min}$
    and $\|\widehat{\bm\gamma}_{\widehat B\cup B} - \widehat{\bm\gamma}^\circ_{\widehat{B}\cup B}\|_2 \geq \| \widehat{\bm\gamma}_{\widehat{A}^*_1\setminus A} - \widehat{\bm\gamma}^\circ_{\widehat{A}^*_1\setminus A}\|_2$. 
    Thus, 
    \begin{equation*}
        \begin{split}
        \lambda_{\min}(n_2^{-1}\widetilde{\bm Z}^\intercal_{\widehat{B}\cup B}\widetilde{\bm Z}_{\widehat{B}\cup B}) \sqrt{|\widehat{A}^*_1\setminus A|}\|\bm\alpha\|_{\min}
        &\leq \frac{2}{n} \sup_{\{S:|S\setminus A|=|\widehat{A}^*_1\setminus A|\}} \|\widehat{\bm e}^\intercal \widetilde{\bm Z}_{S\setminus B}\|_2\\
        &\quad + \frac{3 \sigma_e}{\sqrt{n_2}}  \sqrt{p\lambda_{\max}(\bm Z^\intercal\bm Z)} a\\
        &\leq \frac{2}{n} \sup_{\{S:|S\setminus A|=|\widehat{A}^*_1\setminus A|\}} \|\widehat{\bm e}^\intercal \widetilde{\bm Z}_{S\setminus B}\|_2
        + \sqrt{c_+} \sigma_e \sqrt{\frac{\log(n)}{n}},
        \end{split}
    \end{equation*}
    where the second inequality follows from Condition \ref{condition:sparse-regression} (iv).
    Similarly, on event $\mathcal{E}$, we have $2^{-1}c_{-}\sqrt{|\widehat{A}^*_1\setminus A|}\|\bm\alpha\|_{\min} < 8\sigma_e\sqrt{c_+}(\sqrt{|\widehat{A}^*_1\setminus A|} +1) \sqrt{\log(n)/n}$. 
    
    However, 
    $\|\bm\alpha\|_{\min}\geq 32 \sigma_e \sqrt{c_+} c_{-}^{-1} \sqrt{\log(n)/n}$. 
    This implies that $|\widehat{A}^*_1\setminus A|=0$, and hence $\widehat A = A$ on event $\mathcal{E}$.}
    
    Finally, note that $P(\widehat{A}\neq A)\leq P(\mathcal{E}^c) \leq P(\mathcal{E}_1^c)+P(\mathcal{E}_2^c)$, where 
    the Gaussian tail bounds yields that 
    \begin{equation*}
        \begin{split}
            P(\mathcal{E}_1^c) \leq n^{-3},\quad 
            P(\mathcal{E}_2^c) \leq \sum_{k=1}^{|A|}2\exp( -3 k \log(n) ) \leq 3 n^{-3}.
        \end{split}
    \end{equation*} 
    The proof is completed.
\end{proof}

\begin{proof}[Proof of Theorem 1]
    By Lemma \ref{lemma:sparse-regression}, it suffices to consider the event $\{\widehat{A}=A\}$. Denote the oracle estimator 
    by $(\widehat{\beta}^\circ, \widehat{\bm \alpha}^\circ_A)$, 
    namely the OLS estimator with $A$ known. 
    Then 
    \begin{equation*}
        \begin{pmatrix}
            \widehat{\beta}^{\circ}\\
            \widehat{\bm\alpha}^\circ_A
        \end{pmatrix}
        =
        \begin{pmatrix}
        \widehat{\bm X}^\intercal\widehat{\bm X} & \widehat{\bm X}^\intercal\bm Z_{A}\\
        \bm Z_{A}^\intercal\widehat{\bm X} &\bm Z^\intercal_A\bm Z_A
        \end{pmatrix}^{-1}
        \begin{pmatrix}
            \widehat{\bm X}^\intercal\bm Y\\
            \bm Z_A^\intercal\bm Y
        \end{pmatrix}.
    \end{equation*}
    It follows from matrix algebra that 
    \begin{equation*}
        \begin{split}
            \widehat{\beta}^\circ 
            &= \widehat{\Omega}_X \widehat{\bm X}^\intercal\bm Y - 
            \widehat{\Omega}_X \widehat{\bm X}^\intercal\bm Z_A(\bm Z_A^\intercal\bm Z_A)^{-1}\bm Z_A^\intercal\bm Y,\\
            \widehat{\Omega}_X &= (\widehat{\bm X}^\intercal\widehat{\bm X} 
            - \widehat{\bm X}^\intercal\bm Z_A(\bm Z_A^\intercal\bm Z_A)^{-1}\bm Z_A^\intercal\widehat{\bm X})^{-1}.
        \end{split}
    \end{equation*}
    By Lemma \ref{lemma:sir}, $\sqrt{n_1}(\widehat{\bm\theta}-\bm\theta)\dto \bm \xi$. 
    Then by direct calculation,  
\begin{equation*}
    \begin{split}
        &\sqrt{n_2}(\widehat{\beta}^\circ - \beta) \\
        =&  \sqrt{n_2}\Omega_X^{-1}\bm\theta^\intercal\bm Z^\intercal 
        (\bm I-\bm P_{\bm Z_A})\bm e - \sqrt{r}\beta \Omega_X^{-1} 
        (\bm\theta^\intercal\bm Z^\intercal\bm Z - \bm\theta^\intercal\bm Z^\intercal\bm Z_A(\bm Z_A^\intercal\bm Z_A)^{-1} \bm Z_A^\intercal\bm Z) \bm\xi
    +  o_p(1)\\
    =& \zeta - \eta + o_p(1),
    \end{split}
\end{equation*}
where $\bm P_{\bm Z_A} = \bm Z_A(\bm Z_A^\intercal\bm Z_A)^{-1}\bm Z_A^\intercal$. 
Since two samples are independent, we have $\zeta\independent \eta$.
Finally, note that by Lemma \ref{lemma:sparse-regression}, $P(\widehat{\beta} = |\widehat{\beta}^\circ|) \to 1$. 
Hence, we have $n_2^{1/2}(\widehat{\beta}-\beta) = |n_2^{1/2}\beta + \zeta - \eta| - n^{1/2}_2\beta + o_p(1)$, 
which completes the proof.
\end{proof}

\begin{proof}[Proof of Corollaries 1 and 2]
    The desired results follow immediately from Theorem 1. 
\end{proof}

\begin{proof}[Proof of Theorem \ref{theorem:air}]
    Let 
    $$a = n_1^{-1}\sum_{i=1}^{n_1} (\bm z_{1i}^\intercal\widehat{\bm\theta})^2, \qquad \widetilde{a}=n_1^{-1}\sum_{i=1}^{n_1} (\bm z_{1i}^\intercal{\bm\theta})^2,$$ 
    $$b = n_1^{-1}\sum_{i=1}^{n_1} \widehat{m}(x_{1i})\bm z^\intercal_{1i}\widehat{\bm\theta}, \qquad \widetilde{b} = n_1^{-1}\sum_{i=1}^{n_1} {m}_0(x_{1i})\bm z^\intercal_{1i}{\bm\theta}.$$
    Let $\widehat{\bm \Sigma} = n_1^{-1}\sum_{i=1}^{n_1}\bm z_{1i}\bm z_{1i}^\intercal$ 
    and note that $\|\widehat{\bm \Sigma}-\bm \Sigma\|_2 \leq c_1 \sqrt{p/n_1}$. 
    Let $\widehat C_{zw}=n_1^{-1}\sum_{i=1}^{n_1}\bm z_{1i}^\intercal\bm\theta w_{1i}$ and note that 
    $|\widehat{C}_{zw}|\leq c_3\sqrt{1/n_1}$. 
    Then we have 
    \begin{equation*}
        \begin{split}
            |a-\widetilde{a}| 
            &= |(\widehat{\bm\theta}+\bm\theta)^\intercal\widehat{\bm \Sigma}(\widehat{\bm\theta}-\bm\theta_0)| 
            \leq c_2\sqrt{p/n_1},\\
            |\widetilde{a}-\bm\theta^\intercal\bm \Sigma\bm\theta|
            &= |\bm\theta^\intercal(\widehat{\bm \Sigma}-\bm \Sigma)\bm\theta|
            \leq c_1\sqrt{p/n_1},\\
            |b - \widetilde{b}|
            &= |n_1^{-1}\sum_{i=1}^{n_1} (\widehat{m}(x_{1i})-m(x_{1i}))\bm z_{1i}^\intercal\widehat{\bm\theta} 
            + n_1^{-1} \sum_{i=1}^{n_1} m(x_{1i})\bm z_{1i}^\intercal(\widehat{\bm\theta}-\bm\theta_0)|\\
            &\leq \sqrt{n_1^{-1}\sum_{i=1}^{n_1} (\widehat m(x_{1i}) - m(x_{1i}))^2 }\sqrt{\widehat{\bm\theta}^\intercal\widehat{\bm \Sigma}\widehat{\bm\theta}} + C\sqrt{p/n_1}\\ 
            &\leq c n_1^{-\kappa_2} + C\sqrt{p/n_1}, \\
            |\widetilde{b}- \rho^{-1}\bm\theta^\intercal\bm \Sigma\bm\theta|
            &= |\rho^{-1}\bm\theta^\intercal(\widehat{\bm \Sigma}-\bm \Sigma)\bm\theta + \widehat{C}_{zw}|\leq (c_1 + c_3)\sqrt{p/n_1}.
        \end{split}
    \end{equation*}
    Thus, 
    \begin{equation*}
        |\widehat{\rho} - \rho| \leq \left|\frac{a}{b}-\frac{\widetilde{a}}{\widetilde{b}}\right| + 
        \left|\frac{\widetilde{a}}{\widetilde{b}}-\rho\right|
        \leq c_4\max(\sqrt{p/n_1}, n_1^{-\kappa_2}).
    \end{equation*}
    Taken together, we have 
    \begin{equation*}
        \|\widehat{\phi}-\phi\|_{\infty}\leq
        |\widehat{\rho}-\rho| \| \widehat{m} \|_{\infty} + |\rho|\|\widehat{m}-m\|_{\infty}
        \leq c_5\max(\sqrt{p/n_1}, n_1^{-\min(\kappa_1,\kappa_2)}).
    \end{equation*}
    This completes the proof.
\end{proof}

\end{document}